\documentclass[lettersize,journal]{IEEEtran}
\usepackage{array}
\usepackage[caption=false,font=normalsize,labelfont=sf,textfont=sf]{subfig}
\usepackage{textcomp}
\usepackage{stfloats}
\usepackage{url}
\usepackage{verbatim}
\usepackage{graphicx}
\usepackage{cite}
\usepackage[utf8]{inputenc}
\usepackage{amsmath,amssymb,amsfonts}
\usepackage{amsthm}
\usepackage{algorithm}
\usepackage{algpseudocode}
\usepackage{tablefootnote}
\usepackage{graphicx}
\usepackage{tikz}
\usepackage{float}
\usepackage{xcolor}
\usepackage{multirow}
\usepackage{adjustbox}
\usepackage{stfloats}
\usepackage{url}
\usepackage{balance}
\usepackage{epstopdf}

\usepackage{svg}
\begin{document}

\title{Self-Sustainable Metasurface-Assisted mmWave Indoor Communication System}

\author{Zhenyu~Li,
        Ozan~Alp~Topal,
        {\"O}zlem Tu\u{g}fe Demir,
        Emil Bj{\"o}rnson,
        Cicek Cavdar
        % <-this % stops a space
\thanks{Z. Li, O. A. Topal, E. Björnson, and  C. Cavdar are with the School of Electrical Engineering and Computer Science, KTH Royal Institute of Technology, Stockholm, Sweden (e-mail: \{zhenyuli, oatopal, emilbjo, cavdar\}@kth.se).} 
\thanks{Ö. T. Demir is  with the Department of Electrical and Electronics Engineering, TOBB University of Economics and Technology, Ankara, Türkiye, (e-mail: ozlemtugfedemir@etu.edu.tr).}  
\thanks{ This work has been funded by Celtic-Next project RAI-6Green partly supported by Swedish funding agency Vinnova and by Horizon 2020 Beyond5 Project from the ECSEL Joint Undertaking (JU) under grant agreement No 876124. The JU receives support from the EU Horizon 2020 research and innovation programme and Vinnova in Sweden. \"O. T. Demir was supported by 2232-B International Fellowship for Early Stage Researchers Programme funded by the Scientific and Technological Research Council of T\"urkiye. E. Björnson was supported by the FFL18-0277 grant from the Swedish Foundation for Strategic Research.}
}

% The paper headers
% \markboth{Journal of \LaTeX\ Class Files,~Vol.~14, No.~8, August~2021}%
% {Shell \MakeLowercase{\textit{et al.}}: A Sample Article Using IEEEtran.cls for IEEE Journals}

% \IEEEpubid{0000--0000/00\$00.00~\copyright~2021 IEEE}
% Remember, if you use this you must call \IEEEpubidadjcol in the second
% column for its text to clear the IEEEpubid mark.

\maketitle

\begin{abstract}
In the design of a metasurface-assisted system for indoor environments, it is essential to take into account not only the performance gains and coverage extension provided by the metasurface but also the operating costs brought by its reconfigurability, such as powering and cabling. These costs can present challenges, particularly in indoor dense spaces (IDSs). A self-sustainable metasurface (SSM), which retains reconfigurability unlike a static metasurface (SMS), achieves a lower operating cost than a reconfigurable intelligent surface (RIS) by being self-sustainable through power harvesting. In this paper, in order to find a better trade-off between metasurface gain, coverage, and operating cost, the design and performance of an SSM-assisted indoor mmWave communication system are investigated. We first simplify the design of the SSM-assisted system by considering the use of SSMs in a preset-based manner and the formation of coverage groups by associating SSMs with the closest user equipments (UEs). We propose a two-stage iterative algorithm to maximize the minimum data rate in the system by jointly deciding the association between the UEs and the SSMs, the phase-shifts of the SSMs, and allocating time resources for each UE. The non-convexities that exist in the proposed optimization problem are tackled using the feasible point pursuit successive convex approximation method and the concave-convex procedure. To understand the best scenario for using SSM, the resulting performance is compared with that achieved with RIS and SMS. Our numerical results indicate that SSMs are best utilized in a small environment where self-sustainability is easier to achieve when the budget for operating costs is tight.
\end{abstract}

\begin{IEEEkeywords}
Self-sustainable metasurface, mmWave communication, ray tracing, indoor communication system.
\end{IEEEkeywords}

\section{Introduction}

    \IEEEPARstart{T}{he} increasing demand for wireless connectivity in indoor venues presents a significant challenge due to the dense user equipment (UE) population, which is 1.5 times higher than the average traffic per UE in outdoor urban areas~\cite{ericsson_mobility_report_2023}. Constrained by the dense obstructions within indoor environments, achieving adequate coverage for indoor communication systems can further exacerbate the challenge. We refer to the indoor environment, which has blockage-rich properties and high UE density, as the indoor dense space (IDS)~\cite{topal2022mmwave}. Due to both blockage and the dense population, UE mobility is restricted, leading to relatively static channel conditions in an IDS. Some commonly seen IDSs in our daily lives include aircraft cabins and metro wagons. Benefiting from the high available bandwidth, millimeter wave (mmWave) is considered to be one of the promising solutions for providing high-quality wireless connections to many UEs in indoor environments~\cite{pi2011introduction}. However, considering the high blocking attenuation of the mmWave signals~\cite{rappaport2015wideband}, it could still be challenging to design a mmWave communication system in an IDS with a large coverage area.

    One possible solution to the above problem is to use reflecting metasurfaces~\cite{tan2018enabling,anjinappa2021base,perovic2020channel,yildirim2021hybrid,bjornson2019intelligent}. By phase-shifting the impinging signal before it is reflected, the metasurface can manipulate the beam direction, enabling indirect connectivity for UEs situated in non-line-of-sight areas relative to the transmitter. We refer the gain a UE can achieve from this indirect linkage provided by the metasurface as the metasurface gain. This gain closely depends on the capability of the metasurface to dynamically adjust its inserted phase-shifts according to the target's channel condition, a capability referred to as reconfigurability. However, according to the existing studies, reconfigurability comes with the cost of cabling for power supply and signaling for phase controlling~\cite{10480438}, which we refer to as the operating cost. Given the potential for disruption due to the cabling to the original function of the indoor environment and even the safety risks associated with powering, these operating costs are not always affordable when designing a metasurface-assisted IDS communications system. Thus, there is a trade-off between system coverage, metasurface gain, and operating costs. 

    In our previous work~\cite{li2024mixed}, we evaluated the trade-off between the operating costs and the metasurface gain. The system minimum data rate is maximized by mixedly deploying reconfigurable intelligent surfaces (RISs) and static metasurfaces (SMSs) under a given operating cost budget. Here, RIS denotes the metasurface with reconfigurability requiring a high operating cost, while SMS operates in a static manner with almost no operating costs. The results have suggested that the performance gap resulting from having reconfigurability can be mitigated by densely deploying more SMSs. However, this could still be unaffordable for indoor environments such as IDS, where available space for deploying large numbers of such devices is limited. In pursuit of achieving a more optimal balance between trade-offs, as discussed in~\cite{hu2021robust,cheng2022self,9644606,10348506}, a reconfigurable metasurface with the capability to perform energy harvesting to compensate for the power consumption required by its reconfigurability is investigated. With reconfigurability yet a much lower operating cost compared to RIS, this type of metasurface could potentially promise a better solution within this trade-off. In this paper, we refer to this type of metasurface as the self-sustainable metasurface (SSM). 

    The existing study on SSM predominantly concentrates on enhancing system performance while maintaining a low total system power consumption. \cite{hu2021robust} optimizes the sum rate in an SSM-assisted multi-user multiple-input single-output (MISO) downlink system by jointly designing the beamformer, SSM phase-shifts, and harvesting scheme. In \cite{cheng2022self}, to improve the efficiency of wireless power transfer, an absorb-then-reflect scheme is designed to utilize the SSM in a multi-user system. \cite{9644606} minimizes the transmit power by jointly selecting the beamformer, SSM phase-shifts, and reflection amplitude while satisfying the signal-to-noise ratio at the receiver side and adhering to the energy budget of the SSM. In \cite{10348506}, the authors propose a performance metric named energy-data rate outage probability to describe the performance of an SSM-assisted system. Different harvesting schemes are evaluated in terms of minimizing the energy-data rate outage probability under different SSM placement strategies. Advanced as they are, the necessity of using SSM compared to RIS and SMS still lacks thorough investigation. Moreover, there remains a lack of research providing synthesized evaluations of SSM in terms of its benefits and drawbacks. To address this research gap, this paper introduces an SSM-assisted communication system design within IDS. We propose a novel optimization algorithm aimed at maximizing the minimum data rate by jointly selecting SSM's phase-shifts, SSM-UE associations, and allocating time resources for UEs. Additionally, we investigate both the necessity and feasibility of the SSM-assisted system and perform a comprehensive comparison with RIS-assisted and SMS-assisted systems.

% \hfill mds
 
% \hfill August 26, 2015
    \subsection{Contributions}
        Considering the capability of maintaining a low operating cost while preserving the reconfigurability of the metasurface, SSM can be a better passive alternative to RIS than SMS. In this paper, in terms of finding a better solution from the trade-off among metasurface gain, operating cost, and system coverage, we investigate the performance of an SSM-assisted indoor mmWave communication system. The contributions of this paper are outlined as follows:
        \begin{itemize}
            \item We present a large-scale SSM-assisted system designed with low computational complexity for optimization. In this system, the utilization of the SSM is simplified by contemplating its use by following in a pre-designed pattern and by considering the associations between the SSMs and UEs based on the feasibility of achieving self-sustainability.
            \item We formulate a mixed-integer programming (MIP) optimization problem to maximize the minimum achievable data rate in the system by jointly deciding the SSM-UE association, phase-shifts of the SSMs, and allocating time resources for each UE. To tackle the non-convexity that comes with the optimization problem, the feasible point pursuit successive convex approximation (FPP-SCA) and the concave-convex procedure (CCP) are utilized to convexify the constraints. Moreover, we propose a two-stage iterative algorithm based on the convexified optimization problem. 
            \item We investigate the feasibility and necessity of utilizing SSM by checking the SSM contribution, the SSM-UE association, and the self-sustainability condition for SSMs in different positions.
            \item Extending our previous work~\cite{li2024mixed}, where we investigate the RIS- and SMS-assisted system, in this work we focus on SSM, which is reconfigurable yet requires a much lower operating cost than RIS. In assessing the trade-offs between metasurface gain, operating costs, and system coverage, we conducted a thorough comparison of resulting data rates employing SSM, RIS, and SMS across indoor environments with varying sizes. Through our research, we provide a fresh perspective on better scenarios and methodologies for deploying SSM over other types of metasurfaces.
            %Our findings indicate that SSM can achieve a higher data rate compared to SMS in a small cell where SSM can maintain its self-sustainability more easily and a lower operating cost than RIS. However, limited by the self-sustainable constraint, SSM can only be utilized in a smaller range compared to SMS when the budget for operating costs is tight.
        \end{itemize}
        
    \subsection{Structure}
        The paper is organized as follows. In Section~\ref{sec:systemmodel} we describe the SSM-assisted system and model the self-sustainable constraint of an SSM. In Section~\ref{sec:optimization}, we formulate the SSM-assisted data rate optimization problem and describe the simplifications we use to lower the computational complexity of the formulated optimization problem. Later in Section~\ref{sec:convexification}, we describe the convexification process of the optimization problem and propose an iterative SSM-assisted data rate optimization algorithm that maximizes the minimum achievable data rate. A ray tracing (RT) environment and the design of the coverage group and preset based on the obtained channel conditions are provided in Section~\ref{sec:raytracing}. The numerical results after optimizing the SSM-assisted system are described in Section~\ref{sec:results}. Lastly, we conclude the paper in Section~\ref{sec:conclusion}.

        The mathematical notations used in this paper are summarized as follows. $\mathbb{B}$ denotes the binary domain, and $\mathbb{C}$ denotes the complex domain. $\|\cdot\|$ represents the $l_2$ norm, $\|\cdot\|_F$ represents the Frobenius norm, $|\cdot|$ represents the magnitude of a complex value. Additionally, $(\cdot)^*$ and $(\cdot)^T$ represent the conjugate and transpose operation. Moreover, $\operatorname{diag}(\cdot)$ represents the diagonal matrix extension of a vector, and $\mathbb{E}\{\cdot\}$ represents the expected value of a random variable. Lastly, $\mathbf{1}_M$ and $\mathbf{0}_M$ indicate a $M\times 1$ vector with all its entries equal to 1 and 0 respectively. $[\cdot]_m$ takes the value of the $m$-th entry of the vector.

\section{System model}\label{sec:systemmodel}

    We consider the downlink of a mmWave communication system that consists of one BS that serves multiple UEs located in an IDS. The BS is equipped with $N$ antennas in the form of an $\sqrt{N}\times \sqrt{N}$ uniform planar array (UPA), while each UE is equipped with a single antenna. We denote the UE indices as $k \in \{1,2,\ldots,K\}=\mathcal{K}$ where $K$ is the total number of UEs. To alleviate interference, the UEs are assumed to be served orthogonally in time, where $\tau_k$ is the time portion allocated to UE $k$, and $\sum_{k=1}^{K} \tau_k =1$. Taking into account the hardware regulations, the phase-shifts switching speed of a metasurface is restricted. Thus the allocated time portion is limited in the range of $\tau_k\in[\tau_\text{min},1]$ where $0 < \tau_\text{min}<1$ is the minimum allowable time portion. % In order to maximize the data rate of the serving UE, the phase shifts of the RISs (if possible) and the precoder of the BS are reconfigured at each time slot to jointly enhance the channel condition of the UE.
        
\begin{figure}[tb]
        \centering
        \includegraphics[trim={70mm 10mm 80mm 40mm},clip,width=.9\linewidth]{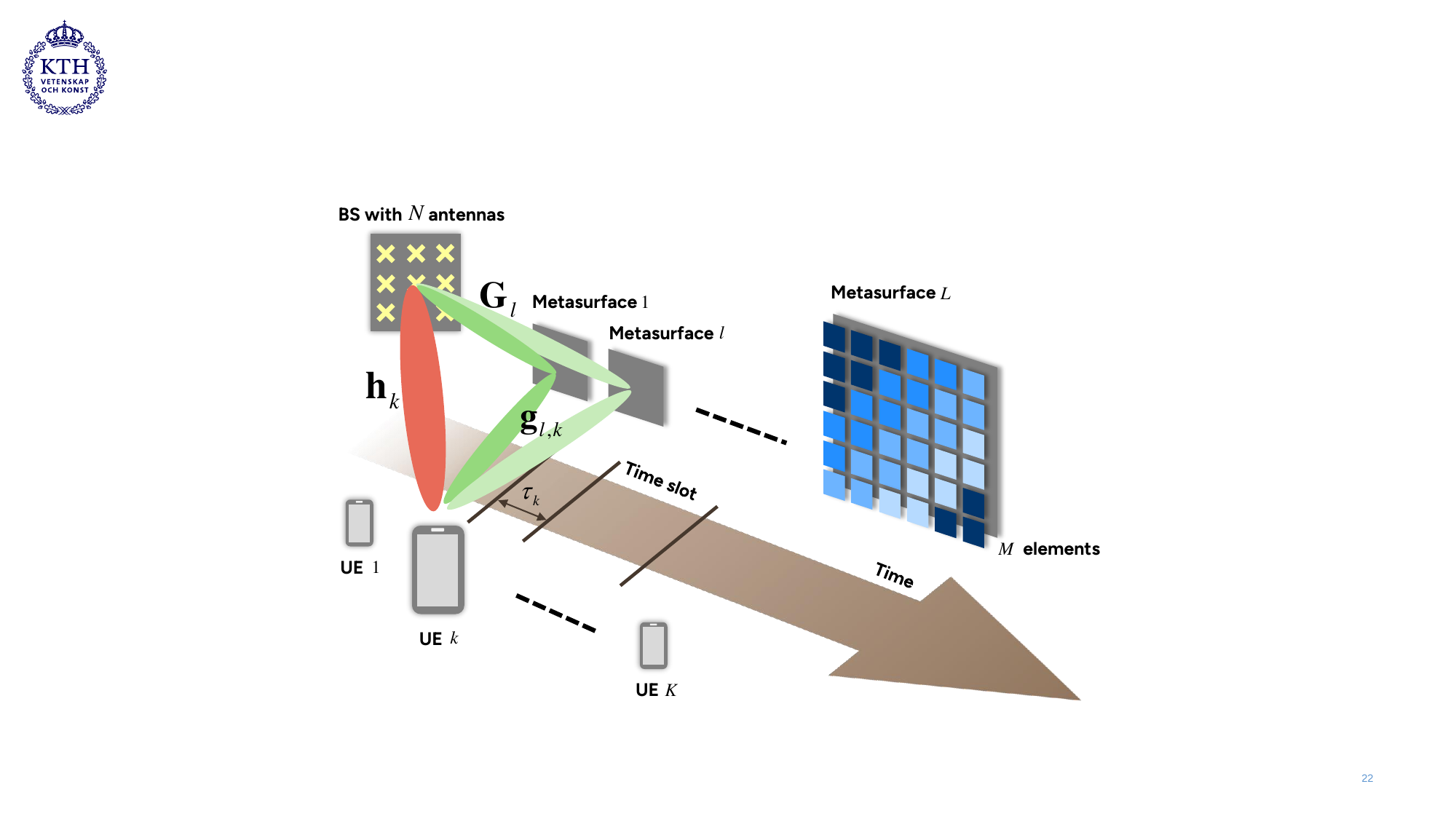}
        \vspace{-4mm}
        \caption{Metasurface-assisted indoor communication system model.}
        \label{fig:systemmodel}
    \end{figure}

\begin{figure}[tb]
        \centering
        \includegraphics[trim={80mm 60mm 80mm 50mm},clip,width=.95\linewidth]{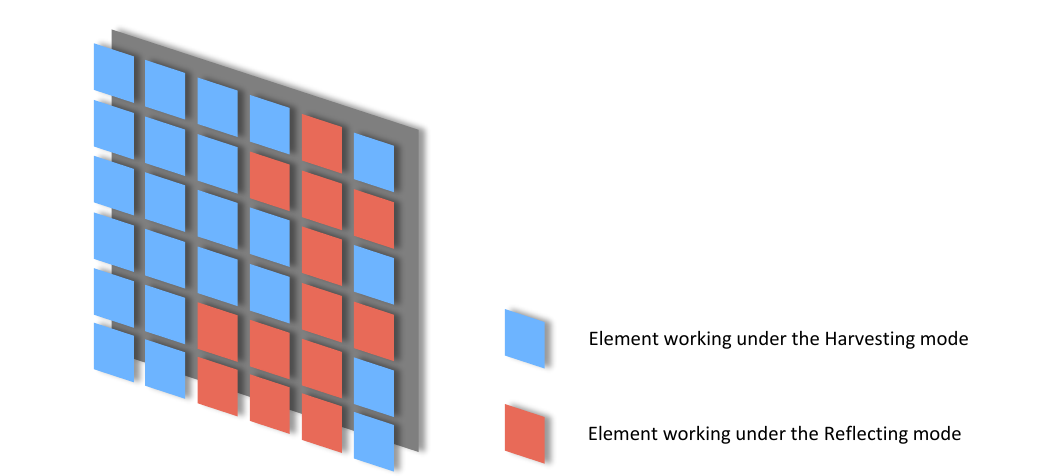}

        \caption{Illustration of the SSM with element splitting scheme.}
        \label{fig:metasurface}
    \end{figure}

    $L$ metasurfaces are deployed in the IDS to improve the downlink rates of the UEs, where each of them has $M$ elements in the form of an $\sqrt{M}\times\sqrt{M}$ UPA. The indices of the elements are denoted as $m\in\{1,2,\ldots,M\}=\mathcal{M}$. The index of an arbitrary surface is denoted as $l \in \{1,2,\ldots, L\}=\mathcal{L}$. Those metasurfaces could be SSMs, but also SMSs or RISs as will be explained in the later analysis. As illustrated in Fig.~\ref{fig:systemmodel},  the direct channel from the BS to UE $k$ is denoted as $\mathbf{h}_k\in\mathbb{C}^N$. Moreover, $\mathbf{G}_{l}\in\mathbb{C}^{M\times N}$ and $\mathbf{g}_{l,k}\in\mathbb{C}^{M}$ denote the channel from the BS to the metasurface $l$ and from the metasurface $l$ to the UE $k$, respectively. Due to the static UEs and geometry dependence of the mmWave links in the IDS, we consider the channels to be deterministic and fixed~\cite{topal2022mmwave}. Later, we obtain realistic channel coefficients for a specific environment by using RT simulations. To achieve self-sustainability, as illustrated in Fig.~\ref{fig:metasurface}, metasurfaces with the element splitting scheme~\cite{10348506} are considered. Each element of the metasurface can select its working mode between \emph{reflecting} and \emph{harvesting}. We define the element working mode indicator vector as $\boldsymbol{\beta}_{l,k}\in\mathbb{B}^M$, the $m$-th entry of the vector, $\beta_{l,k,m}$, indicates the working mode of the $m$-th element of metasurface $l$ during the serving time of UE $k$. $\beta_{l,k,m}=1$ indicates that the $m$-th element of metasurface $l$ is in the reflecting mode when serving UE $k$, and $\beta_{l,k,m}=0$ indicates it is in the harvesting mode. When the reflecting mode is selected for a specific element, the element will be configured to shift the phase of the impinging signal. The power cost of a metasurface element to work under the reflecting mode is denoted as $P^\text{Rf}$. 
    
    When the harvesting mode is selected for an element, the energy of the impinging signal will be harvested via its harvesting circuit. We assume the signals that impinge on the harvesting elements can be fully captured and no signal will be reflected back. The resulting response vector for metasurface $l$ when serving UE $k$ is given as $\boldsymbol{\phi}_{l,k}\in\mathbb{C}^M$
    \begin{equation}
        \boldsymbol{\phi}_{l,k} = \operatorname{diag}(\boldsymbol{\varphi}_{l,k})\boldsymbol{\beta}_{l,k}
    \end{equation}
    % {\color{blue}Let's discuss the above equation.}
    where $\boldsymbol{\varphi}_{l,k}\in\mathbb{C}^M$ is the complex response vector of the phase-shift. In this work, the metasurface reflection loss is neglected, thus $|\varphi_{l,k,m}|=1,\forall l,k,m$ holds, where $\varphi_{l,k,m}$ is the $m$-th entry of $\boldsymbol{\varphi}_{l,k}$.

    
    We aggregate the BS-metasurface-UE channel for mathematical convenience. Defining the aggregated channel from BS via metasurface $l$ to UE $k$ as $\mathbf{H}_{l,k}=\mathbf{G}_l^T\bar{\mathbf{G}}_{l,k}\in\mathbb{C}^{N\times M}$, where $\bar{\mathbf{G}}_{l,k}=\operatorname{diag}(\mathbf{g}_{l,k})$, the received signal at UE $k$ becomes
    \begin{equation}
        y_k = \bigg(\mathbf{h}_k+\sum_{l=1}^L\underbrace{\mathbf{G}_{l}^T\bar{\mathbf{G}}_{l,k}}_{\triangleq \mathbf{H}_{l,k}}\boldsymbol{\phi}_{l,k}\bigg)^T\mathbf{w}_kx_k+n_k,
    \end{equation}
    where $x_k\in\mathbb{C}$ is the transmitted data symbol for UE $k$ and the transmit power is $\mathbb{E}\{|x_k|^2\}=P$, $\forall k$. The unit-norm precoding vector $\mathbf{w}_k\in\mathbb{C}^N$ is selected when serving UE $k$. The independent receiver noise is denoted as $n_k \sim \mathcal{CN}(0,BN_0)$,  where $B$ is the communication bandwidth in Hz and $N_0$ is the noise spectral density in W/Hz.

    Since there is no interference due to the orthogonal scheduling of UEs, a good precoding strategy for the BS is maximum ratio transmission (MRT). The MRT precoder for the overall channel from the BS to UE $k$ is 
    \begin{equation}\label{eq:mrt}
        \mathbf{w}_k = \frac{\left(\mathbf{h}_k+\sum_{l=1}^L\mathbf{H}_{l,k}\boldsymbol{\phi}_{l,k}\right)^*}{\left\|\mathbf{h}_k+\sum_{l=1}^L\mathbf{H}_{l,k}\boldsymbol{\phi}_{l,k}\right\|}.
    \end{equation}
    The resulting signal-to-noise ratio (SNR) of UE $k$ is
    \begin{equation}
        \Gamma_k=\frac{\left\|\mathbf{h}_k+\sum_{l=1}^L\mathbf{H}_{l,k}\boldsymbol{\phi}_{l,k}\right\|^2P}{BN_0},\label{eq:snr}
    \end{equation}
    and
    \begin{equation}
        R_k = \tau_kB\log_2(1+\Gamma_k)\quad \text{bit/s}
    \end{equation}
 is the data rate achieved by UE $k$. 

    \subsection{Self-sustainability of the metasurface}
         In this work, we do not consider any power storage device or centralized power distribution device for the system for minimizing the operating cost purpose. Thus, to achieve self-sustainability, a metasurface should harvest enough power from the impinging radio waves to support its operation. The signals that are received by the harvesting elements of the metasurface $l$ when the system is serving UE $k$ can be expressed as
        \begin{equation}
            \mathbf{y}_{l,k}^\text{Rc} = \operatorname{diag}(\mathbf{1}_{M}-\boldsymbol{\beta}_{l,k})\mathbf{G}_l\mathbf{w}_kx_k
        \end{equation}
        and the power received by the metasurface $l$ when the system is serving UE $k$ is given as
        \begin{equation}
            P^\text{Rc}_{l,k} = \mathbb{E}\left\{\|\mathbf{y}_{l,k}^\text{Rc}\|^2\right\} = \sum_{m=1}^M(1-\beta_{l,k,m})\left|\mathbf{\bar{g}}_{l,m}^T\mathbf{w}_k\right|^2P. \label{eq:powerreceived}
        \end{equation}
        Here, the power received from the thermal noise is neglected. Moreover, $\mathbf{\bar{g}}^T_{l,m} \in \mathbb{C}^{1\times N}$ denotes the $m$-th row of  $\mathbf{G}_{l}$. Note that due to the dependency between the selection of $\boldsymbol{\phi}_{l,k}$ and the beamforming vector $\mathbf{w}_k$, according to \eqref{eq:powerreceived} the received power will also depend on the phase-shifts selection.

        The harvested power is calculated by referring to the non-linear energy harvesting model proposed in~\cite{7999248} as
        \begin{equation}
            P_{l,k}^\text{Hr} = \frac{q_1P_{l,k}^\text{Rc}}{q_2P_{l,k}^\text{Rc}+q_3},
        \end{equation}
        where $q_1$, $q_2$, and $q_3$ are positive constants that are related to the efficiency of the energy harvesting hardware. This model not only accurately captures the non-linear characteristic of a real energy harvesting circuit~\cite{boshkovska2015practical}, but also provide convenience for mathematical processing.

\section{Data rate optimization problem with SSM}\label{sec:optimization}

    For fairness, we aim to maximize the minimum achievable data rate in the IDS. It will be maximized by tuning the phase-shifts of the metasurfaces, selecting working modes for each metasurface, and allocating time resources with respect to each UE. The max-min rate optimization problem is formulated as
    \begin{subequations}
        \begin{align}
            &\textbf{P1}:\operatornamewithlimits{maximize}_{\{\boldsymbol{\varphi}_{l,k},\boldsymbol{\beta}_{l,k},\tau_k\}} \, \operatornamewithlimits{min}_{k\in\mathcal{K}}\, R_k \label{eq:orgobj} \\
            %&\text{subject to}: \notag\\
            &\text{s.t.}\ \sum_{k=1}^K \tau_k \leq 1, \label{eq:timealloc1}\\
            &\hspace{6mm} \tau_k\geq\tau_\text{min}, \quad \forall k,\label{eq:timealloc2}\\
            &\hspace{6mm} |\varphi_{l,k,m}| = 1, \quad \forall l,k,m,\label{eq:lossless}\\
            &\hspace{6mm} \beta_{l,k,m} \in \{0,1\}, \quad \forall l,k,m,\label{eq:binaryprop}\\
            &\hspace{6mm} P_{l,k}^\text{Hr} \geq \sum_{m=1}^M\beta_{l,k,m}P^\text{Rf}, \quad \forall l,k.\label{eq:selfsustain}
        \end{align}
    \end{subequations}

    By maximizing the objective function in~\eqref{eq:orgobj} the minimum achievable data rate among the UEs improved. The constraint in \eqref{eq:timealloc1} makes sure the sum of the allocated time portions does not exceed one and the constraints in \eqref{eq:timealloc2} ensure that the individual allocated time portion is not shorter than the minimum portion $\tau_\text{min}$. The lossless reflection of the metasurface is ensured by the constraints in \eqref{eq:lossless}. The binary working mode selection is guaranteed by the constraints in \eqref{eq:binaryprop}. Finally, the self-sustainability is achieved by fulfilling the constraints in \eqref{eq:selfsustain}. Due to the involvement of the binary variable, $\textbf{P1}$ is formulated as a mixed-integer programming problem.
    
    \subsection{Simplifying the utilization of SSM}

        Despite the non-convexity associated with the constraints in $\textbf{P1}$, the main issue that makes it challenging to solve is the computational complexity. The computational complexity of the mixed-integer programming problem $\textbf{P1}$ increases with the number of integer variables such as $\beta_{l,k,m}$. Given the dense population of the IDS, a large number of integer variables is anticipated, potentially rendering the problem computationally infeasible to solve.  To tackle those issues, in this subsection, we consider the working mode selection of the SSM is determined according to a pre-designed pattern, which we will refer to later as the preset. Moreover, we recast the working mode selection problem into an SSM-UE association problem, which greatly reduces the computational complexity. 
    
        \subsubsection{Preset-based SSM}
            
\begin{figure}[h]
                \centering
                \includegraphics[trim={80mm 38mm 80mm 60mm},clip,width=.95\linewidth]{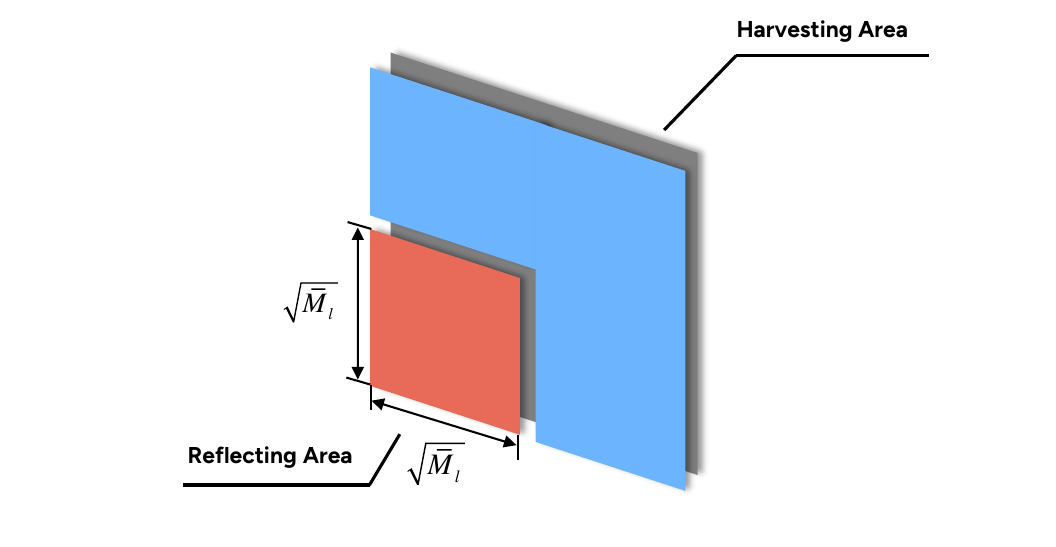}
                \caption{Illustration of the preset design.}
                \label{fig:preset}
            \end{figure}

            For the sake of reducing the number of binary variables in the problem, we consider the mode switching of the SSMs to follow a pre-given $\boldsymbol{\beta}_{l,k}$ for each SSM $l$ when serving different UE $k$, which we refer to as the \emph{preset}. The same preset is considered to be utilized when SSMs are serving different UEs. Thus, $\boldsymbol{\beta}_{l,k}=\boldsymbol{\beta}_l, \forall k$. %we argue that this assumption not only reduces the number of binary variables but also simplyfies the control logic of the SSM. 
            Moreover, for simplicity and ease of analysis, we consider the preset design shown in Fig.~\ref{fig:preset}, where the SSM is divided into the \emph{reflecting area} and the \emph{harvesting area}. The reflecting area of the SSM $l$ consists of $\bar{M}_l$ number of reflecting elements that are of the form of a $\sqrt{\bar{M}_l}\times\sqrt{\bar{M}_l}$ UPA. The remaining elements, which are not in the reflective area, work to harvest power and form the harvesting area. We assume that the reflecting area is always located at the lower left corner of the SSM regardless of their serving target. Notice that only as a simple yet efficient way of using SSM, with this preset-based simplification, optimality of performance is not guaranteed. However, this not only eliminates the binary variable $\beta_{l,k,m}$, which greatly reduces the computational complexity but also makes it easier to compare performance with other types of metasurfaces for later analysis. 
            
           Considering the SSM is used based on a given preset, $\beta_{l,k,m}$ is treated as a constant in the later content, and the equation $\sum_{m=1}^M\beta_{l,k,m} = \bar{M}_l$ holds $\forall l$ and $\forall k$. The sets that consist of the indices of the elements in the reflecting area and in the harvesting area are denoted as $\bar{\mathcal{M}}_l$ and $\hat{\mathcal{M}}_l$, respectively. Those sets satisfy that $\bar{\mathcal{M}}_l\cup\hat{\mathcal{M}}_l = \mathcal{M}$ and $\bar{\mathcal{M}}_l\cap\hat{\mathcal{M}}_l = \varnothing$.

        \subsubsection{SSM-UE association}

            To avoid violating the self-sustainable constraint for a given preset, we introduce a binary SSM-UE association indicator $\alpha_{l,k}\in\mathbb{B}$ to indicate whether SSM $l$ will be assigned to serve UE $k$. Note that the number of  SSM-UE association indicators can still be very high in an IDS communication network. Considering the obstacle-rich property of the IDS and the high attenuation of the mmWave propagation, metasurfaces generally will not be able to offer significant help to a UE that is far away from them. Due to the transmit beamforming, it is also hard for them to capture enough energy to achieve self-sustainability when the system is serving a UE that is in the opposite direction of them. To further reduce the computational complexity, we first manually assign SSMs to UEs based on the coverage of the SSM to form multiple groups, which we refer to as the \emph{coverage group}. Then the SSM-UE association inside the coverage group will be decided jointly while optimizing the minimum data rate. 
        
            We denote the index of the coverage group as $i \in \{1,\ldots, I\}$, where there are $I$ coverage groups in total. Metasurfaces and UEs in group $i$ are indexed as $l \in \mathcal{L}_i\subseteq\mathcal{L}$ and $k \in \mathcal{K}_i\subseteq\mathcal{K} $ respectively. In particular, we point out that the same SSM can be assigned to serve UEs in different coverage groups to make the best use of their coverage. In other words, the intersection of two index sets $\mathcal{L}_i~ \mathcal{L}_j,~\forall i\neq j$ does not necessarily have to be an empty set. Each group $i$ contains $L_i$ metasurfaces and $K_i$ UEs. 
            % {\color{blue}We can clarify whether these sets are overlapping or not.}
            The SNR of UE $k$, which is in coverage group $i$, can be expressed as
            \begin{equation}
                \Gamma_{k}=\frac{\left\|\mathbf{h}_{k}+\sum_{l\in\mathcal{L}_i}\alpha_{l,k}\mathbf{H}_{l,k}\boldsymbol{\phi}_{l,k}\right\|^2P}{BN_0}.
            \end{equation}

    \subsection{Joint data rate optimization algorithm with SSM}
    With the above mechanisms considered, $\textbf{P1}$ is recast as two sub-problems $\textbf{P2-1}$ and $\textbf{P2-2}$. The objective of the sub-problem $\textbf{P2-1}$ is to maximize the minimum achievable data rate in each coverage group by jointly deciding the phase-shifts, SSM-UE association, and allocation of time resources inside the working group. The sub-problem $\textbf{P2-1}$ is given as
            \begin{subequations}
                \begin{align}
                    &\textbf{P2-1}: \operatornamewithlimits{maximize}_{\{\boldsymbol{\varphi}_{l,k},\alpha_{l,k},\tilde{\tau}_{k}\}} \min_{k\in\mathcal{K}_i} \tilde{R}_{k} \ \\
                    &\text{s.t. }\sum_{k\in\mathcal{K}_i}\tilde{\tau}_{k} \leq 1,\\
                    &\hspace{6mm}\tilde{\tau}_{k}\geq\tau_\text{min},\quad \forall k\in\mathcal{K}_i, \\
                    &\hspace{6mm}|\varphi_{l,k,m}|=1, \quad \forall l\in\mathcal{L}_i,k\in\mathcal{K}_i,m,\\
                    &\hspace{6mm}\alpha_{l,k} \in \{0,1\},\quad \forall l\in\mathcal{L}_i,k\in\mathcal{K}_i,\label{eq:11e}\\
                    &\hspace{6mm} P_{l,k}^\text{Hr} \geq \alpha_{l,k}\bar{M}_lP^\text{Rf}, \quad \forall l\in\mathcal{L}_i,k\in\mathcal{K}_i, \label{eq:ssconorg}
                \end{align}
            \end{subequations}
            %{\color{blue}Isn't there any constraint on the summation of $\alpha$'s?}
        where $\tilde{\tau}_{k}$ is the allocated time portion inside the coverage group. Self-sustainability is not a concern for the unassigned SSM, so when $\alpha_{l,k}=0$, the constraint~\eqref{eq:ssconorg} is automatically satisfied. $\tilde{R}_{k}$ is the resulting data rate after allocating the time resources in the coverage group, and is given as
        %{\color{blue}Özlem: I guess the following is not the actual data rate since the allocated time portions will be smaller at the end, we can discuss naming it in the meeting.}
        \begin{align}
            &\tilde{R}_{k}(\boldsymbol{\phi}_{l,k},\alpha_{l,k},\tilde{\tau}_{k}) = \tilde{\tau}_{k}B\log_2\bigg(1+\notag\\
            &\hspace{6mm}\frac{\|\mathbf{h}_{k}+\sum_{l\in\mathcal{L}_i}\alpha_{l,k}\mathbf{H}_{l,k}\boldsymbol{\phi}_{l,k}\|^2P}{BN_0}\bigg),\quad\forall k\in\mathcal{K}_i.   
        \end{align}
        Notice that the binary variable $\alpha_{l,k}$ is coupled with the complex variables $\boldsymbol{\varphi}_{l,k}$ inside the expression of $\tilde{R}_{k}$ which brings non-convexity to the problem. In the same expression, the quadratic form inside the logarithm and the coupling related to $\tilde{\tau}_{k}$ makes the problem non-convex. Besides, in~\eqref{eq:ssconorg} due to the complex form of the MRT precoder and the energy harvesting model, the binary SSM-UE association indicator $\alpha_{l,k}$ and metasurface responses $\boldsymbol{\phi}_{l,k}$ are coupled in a non-convex way. 
        
        In sub-problem $\textbf{P2-2}$, the time resources are allocated in the whole network in terms of maximizing the minimum data rate among the whole system. We denote the output phase-shifts, SSM-UE association, and allocated intra time portion after finding the solution to $\textbf{P2-1}$ for each coverage group $i$ as $\boldsymbol{\phi}_{l,k}^*$, $\alpha_{l,k}^*$, and $\tilde{\tau}_{k}^*$, respectively. We denote the network-wide allocated time portion as $\hat{\tau}_k \in \mathbb{R}$. The resulting achievable data rate for each UE in the network is expressed as
        \begin{equation}
            R_k = \hat{\tau}_{k}\tilde{R}_{k}(\boldsymbol{\phi}_{l,k}^*,\alpha_{l,k}^*,\tilde{\tau}_{k}^*), \quad \forall k.\label{eq:13}
        \end{equation}

        Sub-problem $\textbf{P2-2}$ is given as
        \begin{subequations}
            \begin{align}
                &\textbf{P2-2}: \operatornamewithlimits{maximize}_{\{\hat{\tau}_k\}} \quad \min_{k\in\mathcal{K}} R_k\\
                &\text{s.t. }\hat{\tau}_k\tilde{\tau}_{k}^* \geq \tau_\text{min}, \quad \forall k,\\
                &\hspace{5mm}\sum_{k\in\mathcal{K}}\hat{\tau}_k\tilde{\tau}_{k}^* \leq 1.
            \end{align}
        \end{subequations}
    Except for the non-convexities in $\textbf{P2-1}$, sub-problem $\textbf{P2-2}$ is formatted as a linear programming (LP) problem, which can be easily solved by any convex programming solver. 
    
    In the following section, we will tackle the non-convexities in $\textbf{P2-1}$ and introduce a solvable iterative optimizing algorithm.

        \section{Iterative joint data rate optimization algorithm}\label{sec:convexification}
%\textcolor{blue}{Some part of these operations can be given in Appendix and some part must be here. Let's discuss in the meeting.}
            We introduce the variable $\tilde{R}_{i}^\text{min}$ that represents the minimum data rate after allocating time resources within the coverage group $i$ and recast $\textbf{P2-1}$ as
            \begin{subequations}
                \begin{align}
                    &\textbf{P3-1}: \operatornamewithlimits{maximize}_{\{\boldsymbol{\varphi}_{l,k},\alpha_{l,k},\tilde{\tau}_{k}\}} \tilde{R}_{i}^\text{min} \label{eq:17a}\\
                    &\text{s.t. }\tilde{R}_{k}(\boldsymbol{\phi}_{l,k},\alpha_{l,k},\tilde{\tau}_{k})\geq \tilde{R}_{i}^\text{min}, \quad \forall k\in\mathcal{K}_i, \label{eq:17b} \\
                    &\hspace{5mm}\sum_{k\in\mathcal{K}_i}\tilde{\tau}_{k} \leq 1,\quad\label{eq:15c}\\
                    &\hspace{5mm}\tilde{\tau}_{k}\geq\tau_\text{min},\quad \forall k\in\mathcal{K}_i, \label{eq:15d}\\
                    &\hspace{5mm}|\varphi_{l,k,m}|=1,\quad \forall l\in\mathcal{L}_i,k\in\mathcal{K}_i,m,\label{eq:15e}\\
                    &\hspace{5mm}\alpha_{l,k} \in \{0,1\},\quad \forall l\in\mathcal{L}_i,k\in\mathcal{K}_i,\\
                    &\hspace{5mm} P_{l,k}^\text{Hr} \geq \alpha_{l,k}\bar{M}_lP^\text{Rf}, \label{eq:17h} \quad \forall l\in\mathcal{L}_i,k\in\mathcal{K}_i.
                \end{align}
            \end{subequations}
            
            To remove the coupling between the binary preset selector $\alpha_{l,k}$ and the complex continuous phase-shifts $\boldsymbol{\phi}_{l,k}$, we introduce an auxiliary variable $\mathbf{z}_{l,k}\in\mathbb{C}^{M}$ to replace the term $\alpha_{l,k}\boldsymbol{\phi}_{l,k}$ with following constraint
            \begin{align}
                \left|\left[\mathbf{z}_{l,k}\right]_m\right| \leq \alpha_{l,k},\quad \forall l\in\mathcal{L}_i,k\in\mathcal{K}_i,m. \label{eq:16}
            \end{align}
            When $\alpha_{l,k} = 1$, each entry of $\mathbf{z}_{l,k}$ will be bounded inside the unit circle. The optimal solution is observed to be found on the unit circle, thus fulfilling the constraint~\eqref{eq:15e}. When $\alpha_{l,k}=0$, $\mathbf{z}_{l,k}$ will be forced to be $\mathbf{0}_M$, thus canceling the contribution from SSM $l$ to UE $k$.
            
            Next, we resolve the non-convexities in \eqref{eq:17b}. We replace $\tilde{R}_i^\text{min}$ with $\tilde{r}_i^\text{min}$ in \eqref{eq:17a} and \eqref{eq:17b}, where $\tilde{r}_i^\text{min}=\sqrt{\tilde{R}_i^\text{min}}$. This will enable the construction of a convex form below. Moreover, since the square root is a monotonically increasing function, this replacement will not change the solution to the optimization problem. Next, we introduce the auxiliary variable $e_{k}\in\mathbb{R}$, and \eqref{eq:17b} can be written as
            \begin{equation}
                e_{k} \geq 2^{\frac{\left(\tilde{r}_{i}^\text{min}\right)^2}{B\tilde{\tau}_{k}}}, \quad \forall k\in\mathcal{K}_i, \label{eq:19}
            \end{equation}
            \begin{equation}
                1+\frac{\left\|\mathbf{h}_{k}+\sum_{l\in\mathcal{L}_i}\mathbf{H}_{l,k}\mathbf{z}_{l,k}\right\|^2P}{BN_0}\geq e_{k}, \quad \forall k\in\mathcal{K}_i. \label{eq:20}
            \end{equation}
            Furthermore, we introduce another auxiliary variable $f_{k}\in\mathbb{R}$ in place of the quadratic-over-linear term that appears at the right side of \eqref{eq:19} as 
            % \textcolor{blue}{Ozan: I am conflicted here. We have too many auxiliary variables that we actually do not need. Let's discuss on this. }
            \begin{equation}
                e_{k} \geq 2^{f_{k}}, \quad \forall k\in\mathcal{K}_i, \label{eq:21}
            \end{equation}
            \begin{equation}
                f_{k} \geq \frac{\left(\tilde{r}^\text{min}_i\right)^2}{B\tilde{\tau}_{k}},\quad \forall k\in\mathcal{K}_i, \label{eq:22}
            \end{equation}
            where \eqref{eq:21} is in the form of an exponential cone constraint and \eqref{eq:22} can be further formatted as a second-order cone (SOC) constraint as
            \begin{equation}
                f_{k}+\tilde{\tau}_{k} \geq \left\|\begin{bmatrix} \sqrt{2/B}\tilde{r}^\text{min}_i & f_{k} & \tilde{\tau}_{k}\end{bmatrix}\right\|, \quad \forall k\in\mathcal{K}_i. \label{eq:21a}
            \end{equation}
            With the above processes, the coupling related to $\tilde{\tau}_{k}$ has now been removed. However with the quadratic form appearing at the left side of the inequality in \eqref{eq:20}, the constraint is still not convex. To deal with this issue, we first expand the norm-square term as
             \begin{align}
                &\mathbf{z}_{k}^H\underbrace{\mathbf{H}_{k}^H\mathbf{H}_{k}}_{\triangleq \mathbf{A}_{k}}\mathbf{z}_{k}+2\Re\left(\mathbf{z}_{k}^H\underbrace{\mathbf{H}_{k}^H\mathbf{h}_{k}}_{\triangleq \mathbf{b}_{k}}\right)+\underbrace{\mathbf{h}_{k}^H\mathbf{h}_{k}}_{\triangleq c_{k}} \notag \\
                &\hspace{4mm}\geq \frac{BN_0(e_{k}-1)}{P}, \quad \forall k\in\mathcal{K}_i,\label{eq:normexpand}
            \end{align}
           % {\color{blue}Özlem: We need a different indexing here. Let's discuss this.}
            where $\mathbf{H}_{k} = [\mathbf{H}_{l_{i,1},k},\ldots,\mathbf{H}_{l_{i,L_i},k}] \in \mathbb{C}^{N\times L_iM}$ and $\mathbf{z}_{k}=[\mathbf{z}^T_{l_{i,1},k},\ldots,\mathbf{z}^T_{l_{i,L_i},k}]^T\in\mathbb{C}^{L_iM}$. Here, $l_{i,j}$ denotes the index of the $j$th element in $\mathcal{L}_i$. Next, we utilize the FPP-SCA algorithm~\cite{6954488} to relax \eqref{eq:normexpand}. With $\mathbf{A}_{k}$ as a postive semi-definite matrix, for any arbitrary vector $\boldsymbol{\zeta}_{k}$ it holds that
            \begin{equation}
                \mathbf{z}_{k}^H(-\mathbf{A}_{k})\mathbf{z}_{k}\leq2\Re\left(\boldsymbol{\zeta}_{k}^H(-\mathbf{A}_{k})\mathbf{z}_{k}\right)-\boldsymbol{\zeta}_{k}^H(-\mathbf{A}_{k})\boldsymbol{\zeta}_{k}, \forall k\in\mathcal{K}_i.
            \end{equation}
            At each iteration $\epsilon$, by inserting the non-negative slack variable $s_{k}$ and replacing $\boldsymbol{\zeta}_{k}$ with the previously obtained solution $\mathbf{z}_{k}^{(\epsilon-1)}$, we can replace the quadratic term in \eqref{eq:20} with its affine approximation as 
             \begin{align}
                &\hspace{-2mm}-2\Re\left(\left(\mathbf{z}_{k}^{(\epsilon-1)}\right)^H\mathbf{A}_{k}\mathbf{z}_{k}\right)+\left(\mathbf{z}_{k}^{(\epsilon-1)}\right)^H\mathbf{A}_{k}\mathbf{z}_{k}^{(\epsilon-1)}\notag\\
                &\hspace{1mm}-2\Re\left(\mathbf{z}_{k}^H\mathbf{b}_{k}\right)\leq s_{k}+c_{k}+\frac{BN_0}{P}(1-e_{k}), \quad \forall k\in\mathcal{K}_i,\label{eq:fppsca2}
            \end{align}
            \begin{equation}
                s_{k} \geq 0, \quad \forall k\in\mathcal{K}_i.\label{eq:slack1}
            \end{equation}

            Next, we tackle the non-convexity in the self-sustainable constraint. Introducing an auxiliary variable $u_{l,k}$ and making it larger than or equal to the consumed power but less than or equal to the harvested power, the inequality~\eqref{eq:17h} is equivalent to the following constraints:
            \begin{equation}
                \alpha_{l,k}\bar{M}_lP^\text{Rf} \leq u_{l,k}, \quad \forall l\in\mathcal{L}_i,k\in\mathcal{K}_i,\label{eq:26}
            \end{equation}
            \begin{align}
                &\frac{q_1\sum_{m=1}^M(1-\beta_{l,k,m})|\bar{\mathbf{g}}_{l,m}^T\mathbf{w}_{k}|^2P}{q_2\sum_{m=1}^M(1-\beta_{l,k,m})|\bar{\mathbf{g}}_{l,m}^T\mathbf{w}_{k}|^2P+q_3} \notag\\
                &\hspace{3mm} \geq u_{l,k},\quad \forall l\in\mathcal{L}_i,k\in\mathcal{K}_i. \label{eq:29}
            \end{align}
            Notice that $\mathbf{w}_{k}$ is in a non-convex functional relationship with $\mathbf{z}_{l,k}$, thus~\eqref{eq:29} is not convex. To tackle this, we multiply both the denominator and numerator of the left side of~\eqref{eq:29} by the term $\|\mathbf{h}_{k}+\mathbf{H}_{k}\mathbf{z}_{k}\|^2$. We further introduce auxiliary variables $t_{l,k,m}\in\mathbb{R}$ and $d_{k}\in\mathbb{R}$ with following constraints:
            \begin{align}
                & |\bar{\mathbf{g}}_{l,m}^H(\mathbf{h}_{k}+\mathbf{H}_{k}\mathbf{z}_{k})|^2P \geq t_{l,k,m}, \quad \forall l\in\mathcal{L}_i,k\in\mathcal{K}_i,m, \label{eq:28}\\
                & \|\mathbf{h}_{k}+\mathbf{H}_{k}\mathbf{z}_{k}\|^2\leq d_{k}, \quad \forall k\in\mathcal{K}_i,\label{eq:29a}\\
                & t_{l,k,m} \leq \mathcal{C}(1-\beta_{l,k,m}), \quad \forall l\in\mathcal{L}_i,k\in\mathcal{K}_i,m,\label{eq:32}
            \end{align}
            where $\mathcal{C}$ is a sufficiently large positive number. When $\beta_{l,k,m}=1$, the constraint~\eqref{eq:32}  will force $t_{l,k,m}=0$, thus avoiding that any reflecting element contributes to the energy harvesting. When $\beta_{l,k,m}=0$, the large constant $\mathcal{C}$ makes sure that $t_{l,k,m}$ will not be bounded by~\eqref{eq:32}. With all the newly introduced auxiliary variables injected, \eqref{eq:29} can be expressed as
            \begin{equation}
                \frac{q_1\sum_{m=1}^Mt_{l,k,m}}{q_2\sum_{m=1}^Mt_{l,k,m}+q_3d_{k}} \geq u_{l,k},\quad \forall l\in\mathcal{L}_i,k\in\mathcal{K}_i. \label{eq:31}
            \end{equation}
            However, after replacement, there are still constraints expressed in a non-convex way. The quadratic term appears at the left side of the inequality constraint \eqref{eq:28} still brings non-convexity, and the fractional term at the left side of the inequality constraint \eqref{eq:31} also introduces non-convexity. 

            We first handle the non-convexity in \eqref{eq:28}. Similar to \eqref{eq:20}, where the inequality is formulated as a quadratic term greater than or equal to an affine term, \eqref{eq:28} can also be convexified using FPP-SCA. We first expand the quadratic term as
            \begin{align}
                &\underbrace{\mathbf{h}_{k}^H\mathbf{\bar{g}}_{l,m}\mathbf{\bar{g}}_{l,m}^H\mathbf{h}_{k}}_{\triangleq \hat{c}_{l,k,m}}+2\Re\left(\mathbf{z}_{k}^H\underbrace{\mathbf{H}_{k}^H\mathbf{\bar{g}}_{l,m}\mathbf{\bar{g}}_{l,m}^H\mathbf{h}_{k}}_{\triangleq \hat{\mathbf{b}}_{l,k,m}}\right) \notag\\
                &+\mathbf{z}_{k}^H\underbrace{\mathbf{H}_{k}^H\mathbf{\bar{g}}_{l,m}\mathbf{\bar{g}}_{l,m}^H\mathbf{H}_{k}}_{\hat{\mathbf{A}}_{l,k,m}}\mathbf{z}_{k} \geq \frac{t_{l,k,m}}{P},\quad \forall l\in\mathcal{L}_i,k\in\mathcal{K}_i,m.
            \end{align}
            Then \eqref{eq:28} can be approximated as an affine inequality with the non-negative slack variable $s_{l_i,k_i,m}'$ as 
            \begin{align}
                &-2\Re\left(\left(\mathbf{z}_{k}^{(\epsilon-1)}\right)^H\hat{\mathbf{A}}_{l,k,m}\mathbf{z}_{k}\right)\notag\\
                &\hspace{2mm}+\left(\mathbf{z}_{k}^{(\epsilon-1)}\right)^H\hat{\mathbf{A}}_{l,k,m}\mathbf{z}_{k}^{(\epsilon-1)}-2\Re\left(\mathbf{z}_{k}^H\hat{\mathbf{b}}_{l,k,m}\right)\leq s_{l,k,m}'\notag\\
                &\hspace{2mm}+\hat{c}_{l,k,m}-\frac{t_{l,k,m}}{P}, \quad \forall l\in \mathcal{L}_i,k\in \mathcal{K}_i,m \label{eq:33a}
            \end{align}
            \begin{equation}
                s_{l,k,m}' \geq 0,\quad \forall l\in\mathcal{L}_i,k\in\mathcal{K}_i,m. \label{eq:34a}
            \end{equation}

            Then we tackle the non-convexity in \eqref{eq:31}. We reformulate \eqref{eq:31} as
            \begin{align}
                &\left(u_{l,k}+q_2\sum_{m=1}^Mt_{l,k,m}+q_3d_{k}\right)^2\leq2q_1\sum_{m=1}^Mt_{l,k,m}\notag \\&+u_{l,k}^2+\left(q_2\sum_{m=1}^Mt_{l,k,m}+q_3d_{k}\right)^2,\quad \forall l\in\mathcal{L}_i,k\in\mathcal{K}_i,m.\label{eq:35}
            \end{align}
            We introduce the auxiliary variable $v_{l,k}\in\mathbb{R}$ and let
            \begin{equation}
                v_{l,k}=q_2\sum_{m=1}^Mt_{l,k,m}+q_3d_{k}\quad\forall l\in\mathcal{L}_i,k\in\mathcal{K}_i.
            \end{equation}
            Then \eqref{eq:35} can be expressed with the help of $v_{l,k}$ as
            \begin{align}
                (u_{l,k}+v_{l,k})^2\leq 2q_1\sum_{m=1}^Mt_{l,k,m}+u_{l,k}^2+v_{l,k}^2\quad l\in\mathcal{L}_i,k\in\mathcal{K}_i.
            \end{align}
            The quadratic terms that appear at the right side of the inequality can be convexified by utilizing the concave-convex procedure (CCP)~\cite{6788812}. Replacing the $u_{l,k}^2$ and $v_{l,k}^2$ with their first-order Taylor expansion at their optimal solution obtained in the previous iteration as
            \begin{align}
                &(u_{l,k}+v_{l,k})^2\leq2q_1\sum_{m=1}^Mt_{l,k,m}  \label{eq:38}\\
                &+ \left(u_{l,k}^{(\epsilon-1)}\right)^2+2\left(u_{l,k}^{(\epsilon-1)}\right)\left(u_{l,k}-u_{l,k}^{(\epsilon-1)}\right)\notag \\
                &+\left(v_{l,k}^{(\epsilon-1)}\right)^2+2\left(v_{l,k}^{(\epsilon-1)}\right)\left(v_{l,k}-v_{l,k}^{(\epsilon-1)}\right)\notag +s_{l,k}'',\notag \\
                &\forall l\in\mathcal{L}_i,k\in\mathcal{K}_i\notag
            \end{align}
            \begin{equation}
                s_{l,k}''\geq 0,\quad \forall l\in\mathcal{L}_i,k\in\mathcal{K}_i, \label{eq:39}
            \end{equation}
            where $s_{l_i,k_i}''$ is the introduced non-negative slack variable for taking the affine approximation based on CCP. 

            All the non-convexities except for the binary constraint~\eqref{eq:11e} in $\textbf{P3-1}$ have now been tackled. The~\eqref{eq:11e} can be handled by MIP solver MOSEK~\cite{mosek_2024} with the branch-and-bound method. Together, the solvable optimization problem can be formulated as 
            \begin{align}
                &\textbf{P4-1}: \operatornamewithlimits{maximize}_{\left\{\begin{tabular}{c c c}
                     $\mathbf{z}_{l,k}$ & $\alpha_{l,k}$ & $\tilde{\tau}_{k}$\\
                     $e_{k}$ & $f_{k}$ & $u_{l,k}$ \\
                     $t_{l,k,m}$ &  $d_{k}$ & $v_{l,k}$ \\
                     $s_{k}$ & $s_{l,k,m}'$ & $s_{l,k}''$
                \end{tabular}\right\}} \tilde{R}_{i}^\text{min}-\bar{s}_i\\
                & \text{s.t. }\eqref{eq:11e},\eqref{eq:15c},\eqref{eq:15d},\eqref{eq:16},\eqref{eq:21},\eqref{eq:21a},\eqref{eq:fppsca2},\eqref{eq:slack1}, \notag\\
                &\hspace{5.5mm}\eqref{eq:26},\eqref{eq:29a},\eqref{eq:32},\eqref{eq:33a},\eqref{eq:34a},\eqref{eq:38},\eqref{eq:39},\notag
            \end{align}
            where $\bar{s}_i$ is the penalty term for introducing the slack variables after utilizing approximation of some terms when casting $\textbf{P4-1}$ over coverage group $i$. The value of the non-negative slack variables is positively related to the gap between the approximation and the original term. By taking penalties for the slack variables, the algorithm will push to find an optimal solution that is close to the optimal solution of the original problem. Defining the  $\omega$, $\omega'$, and $\omega''$ as the positive penalty coefficients for $s_{k}$, $s_{l,k,m}'$, and $s_{l,k}''$ respectively,  $\bar{s}_i$ is defined as
            \begin{align}
                \bar{s}_i =& \omega\sum_{k\in\mathcal{K}_i}s_{k}+\omega'\sum_{l\in\mathcal{L}_i}\sum_{k\in\mathcal{K}_i}\sum_{m=1}^Ms_{l,k,m}'\notag\\
                &+\omega''\sum_{l\in\mathcal{L}_i}\sum_{k\in\mathcal{K}_i}s_{l,k}''.
            \end{align}
            \begin{figure}[tb]
                \centering
                \includegraphics[trim={57mm 35mm 75mm 41mm},clip,width=.95\linewidth]{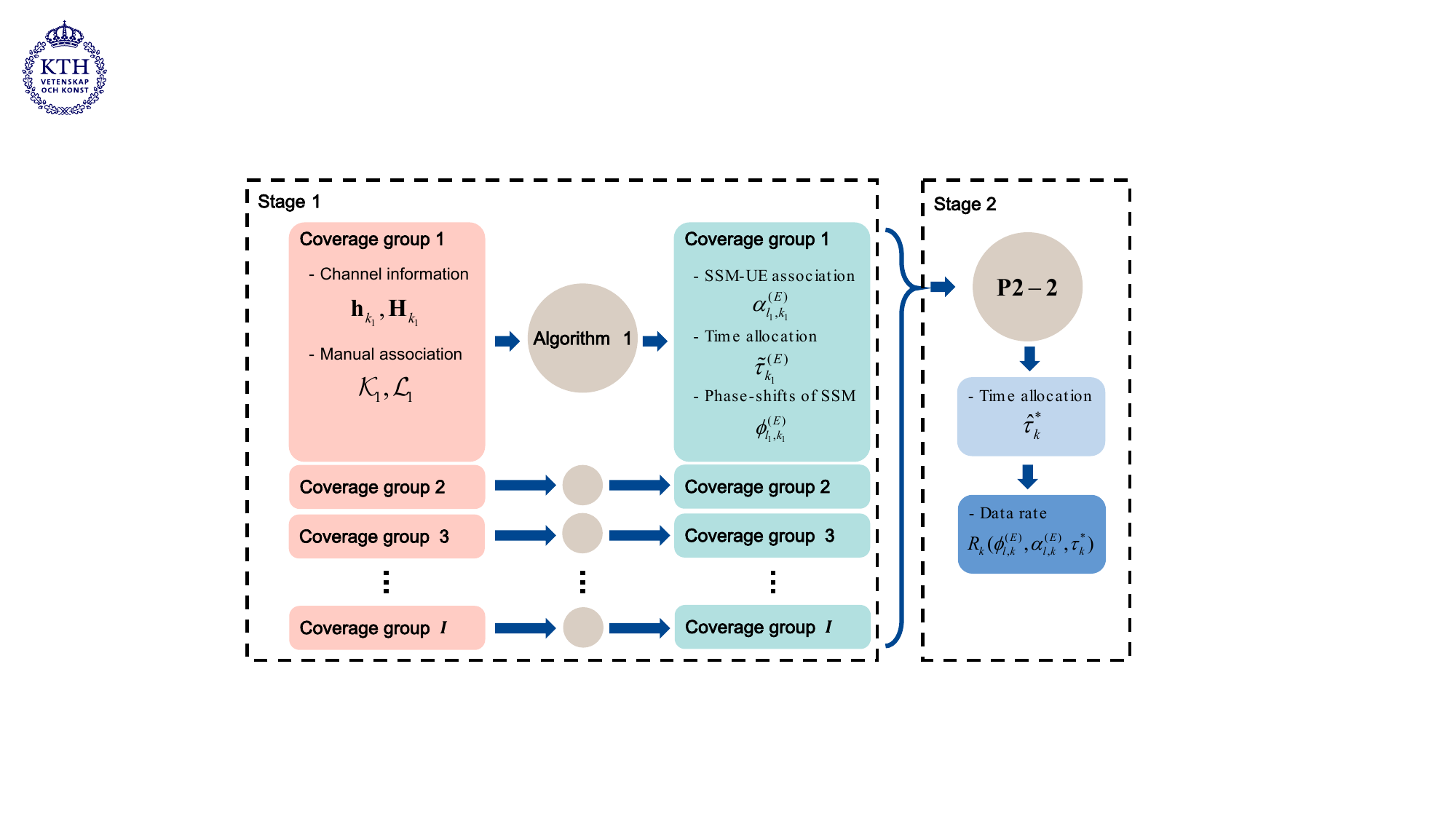}
                \caption{Flowchart of the two-stage iterative  data rate optimization algorithm.}
                \label{fig:flowchart}
            \end{figure}
            The iterative joint data rate optimization algorithm is proposed as a two-stage process. The flowchart of the two-stage optimization algorithm is illustrated in Fig.~\ref{fig:flowchart}. In the first stage, the whole communication system is divided into coverage groups manually. The index sets of the consisted SSMs and UEs $\mathcal{L}_i$ and $\mathcal{K}_i$ are thereby obtained. Then by inputting the obtained index sets along with the channel information associated with the coverage group, Algorithm~\ref{alg:intraoptimize} is utilized for all the formed coverage groups of the system. Using the outputs from the Algorithm~\ref{alg:intraoptimize} with respect to all the coverage groups, optimized $\tilde{R}_{k}, \forall k\in\mathcal{K}$ can be obtained. 
            In the second stage, by inserting $\tilde{R}_{k}, \tilde{\tau}_{k}^{(E)}, \forall k\in\mathcal{K}$ into $\textbf{P2-2}$, and solving the optimization problem, the time resources will be allocated network-wide to maximize the minimum data rate in the whole system. 
            
            \begin{algorithm}[H]
                \caption{Joint coverage group minimum data rate maximization algorithm} \label{alg:intraoptimize}
                \begin{algorithmic}[1]
                    \State {\bf Given:} The manual SSM-UE association made to coverage group $\mathcal{K}_i$, $\mathcal{L}_i$. The channel information of the coverage group $\mathbf{h}_{k}$, $\mathbf{H}_{k}$ \State {\bf Initialization:} Initialize $\mathbf{z}_{k}^{(0)}$ randomly while keeping $|\mathbf{z}_{l,k,m}^{(0)}|=1, ~ \forall l,k,m$. Set the iteration counter to $\epsilon=0$. Set the maximum iteration number to $E$. 
                    \While {$\epsilon < E$}
                        \State $\epsilon \gets \epsilon+1$
                        \State Solve \textbf{P4-1}  and set $\mathbf{z}_{k}^{(\epsilon)},~\forall k\in\mathcal{K}_i$ to its solution
                         
                    \EndWhile
                     %\State $z_{l,k,m}^{(E)} \gets z_{l,k,m}^{(E)}\Big/\vert z_{l,k,m}^{(E)}\vert, \quad \forall l, k, m$
                    \State{\textbf{Output:}} The SSM-UE association $\alpha_{l,k}^{(E)}$, phase-shift configurations $\boldsymbol{\phi}_{l,k}^{(E)}$ of SSMs, and allocated time portions $\tilde{\tau}_{k}^{(E)}$
                \end{algorithmic}
            \end{algorithm}
            
\section{Simulation design}\label{sec:raytracing}
    Considering the static property of the IDS channel and the complexity of the IDS environment, RT simulations are particularly well suited to capture realistic channel conditions. In this paper, the RT simulation is performed using the commercial RT platform Wireless Insite~\cite{Remcom}.
            
\begin{figure}[tb]
        
            \centering
            \subfloat[]{
                \label{fig:geoA}
                \includegraphics[trim={80mm 20mm 105mm 50mm},clip,width=0.4\linewidth]{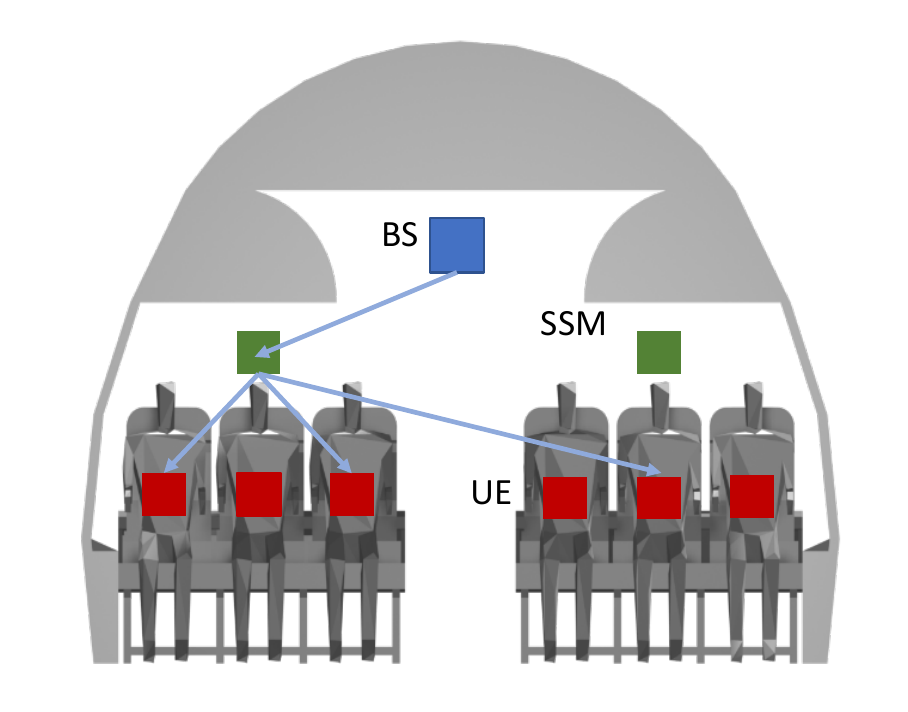}             \vspace{-8mm}
            }
            \subfloat[]{
                \label{fig:geoC}
                \includegraphics[trim={35mm 10mm 157mm 53mm},clip,width=0.5\linewidth]{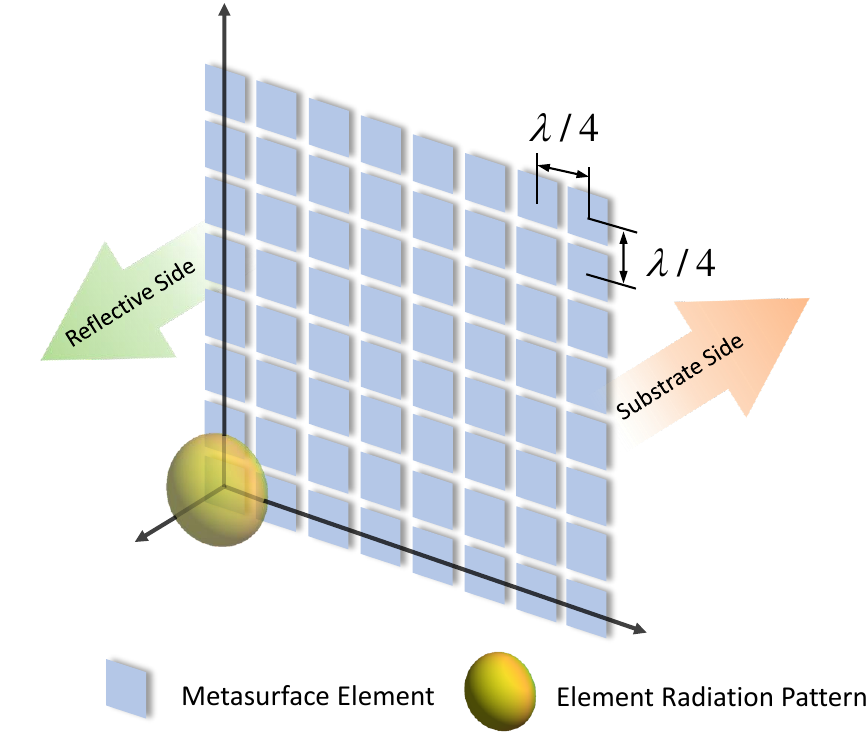}             \vspace{-8mm}
            }

            \subfloat[]{
                \label{fig:geoB}
                \includegraphics[trim={40mm 30mm 65mm 30mm},clip,width=0.93\linewidth]{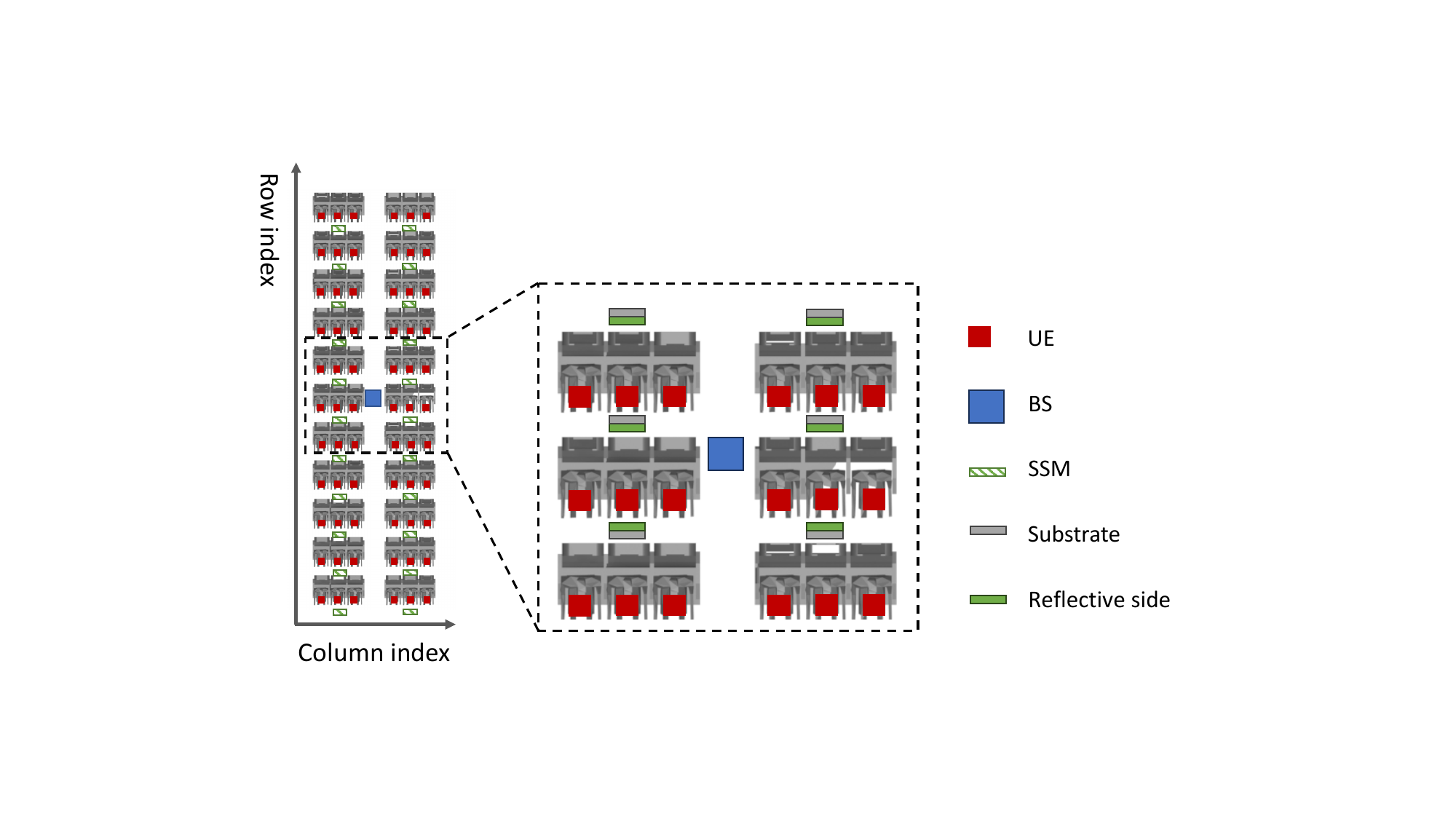}
            }

            \caption{Illustration of the metasurface, BS, and UE placement in the IDS. (a) The front view of the cabin; (b) the detailed metasurface structure, (c) the device layout in the cabin.}
            \label{fig:geometry}
        \end{figure}

    As an example of an IDS, we consider an aircraft cabin with 31 rows and 186 seats full of passengers. Fig.~\ref{fig:geometry} shows the considered aircraft cabin environment. The BS is equipped with an $8\times 8$ UPA of isotropic antennas and is placed close to the ceiling in the middle of the cabin. The antenna array surface of the BS is parallel to the cabin floor. We use a single isotropic antenna to represent the UE held by the passenger. All UEs are set to be higher than the seats to simulate the case when passengers are sitting in their seats and using their cell phones to require service from the BS. The detailed geometry condition of the considered environment can be found in our previous work~\cite{li2023mmwave}.
    
        Metasurfaces are not provided as a module by the RT simulator. Hence, we modeled them as antenna sets in the simulator and synthesized the cascaded channel. The detailed metasurface structure we modeled is given in Fig.~\ref{fig:geoC}. As indicated in~\cite{di2020smart}, the radiation pattern of the metasurface element is similar to that of a cosine antenna. Therefore, we use an $32\times 32$ transceiver cosine antenna array to model the metasurface where each cosine antenna represents one metasurface element. The metasurface elements are placed on an impenetrable substrate that is modeled as a perfect wave absorber. Correspondingly, we align the direction in which the antenna gain of each cosine antenna reaches its maximum perpendicular to the array surface. As shown in Fig.~\ref{fig:geoC}, the cosine antenna only serves half of the sphere, which we will refer to as the reflecting side in this paper. The other half sphere where the antenna gain is zero mimics the wave-absorbing effect of the substrate. After the RT simulation, the channel impulse responses (CIRs) obtained from the BS-metasurface link and the metasurface-UE link are regarded as $\mathbf{G}_l$ and $\mathbf{g}_{l,k}$ respectively.
        
        In our previous work, we observed that the highest SNR to the UEs are obtained when the metasurfaces are placed above the middle seats on both sides of the corridor in each row, with the reflective side perpendicular to the ground and facing the BS ~\cite{li2023mmwave}. With this deployment, all metasurfaces will be line-of-sight to the BS. With two SSMs deployed in each row, there are $L=62$ SSMs deployed in the cabin. Fig.~\ref{fig:geometry} illustrates the geometric placement of the UEs, metasurfaces, and BS in the cabin, where the material details of the airplane cabin can be found in our previous work~\cite{li2023mmwave}. %Considering the huge attenuation along the propagation of the mmWave, only the signals that propagate in a direct path or within two rounds of reflections will be considered during the RT simulation.
        The detailed simulation settings are listed in Table~\ref{tab:simulationparameters}.

        \begin{table}[tb]
            \centering
            \caption{Simulation parameters for the RT.}
            \label{tab:simulationparameters}
            \vspace{-2mm}
            \begin{tabular}{|l|c|}
            \hline
            Carrier frequency      & $28$\,GHz         \\ \hline
            Bandwidth              & $1$\,GHz          \\ \hline
            BS antenna type        & Isotropic      \\ \hline
            Polarization           & V - V          \\ \hline
            Transmit power               & $30$\,dBm          \\ \hline
            Number of BS antennas        & $64 $\\ \hline
            BS antenna spacing    & $\lambda / 2$  \\ \hline
            Metasurface element spacing   & $\lambda / 4$  \\ \hline
            \end{tabular}
    
        \end{table}
    
    \subsection{Coverage group design}

        \begin{figure}[tb]
                \centering
                \includegraphics[width=.95\linewidth]{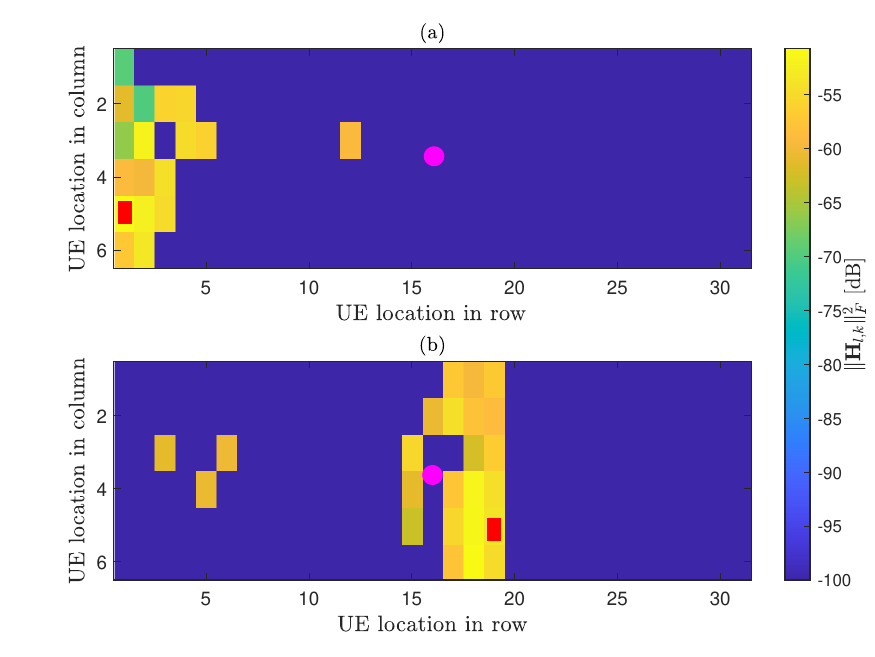}
                \caption{The cascaded BS-SSM-UE channel gain with respect to a SSM that is (a) far-away (b) near to the BS. The red rectangular represents the checked SSM, and the pink circle represents the BS.}
                \label{fig:solometasurface}
        \end{figure}
        
\begin{figure}[t]
            \centering
            \includegraphics[trim={110mm 70mm 85mm 70mm},clip,width=0.93\linewidth]{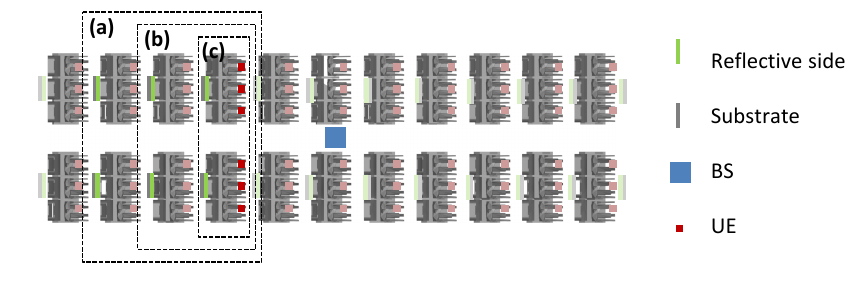}
            \caption{Illustration of different coverage group designs. Coverage group consists of (a) 6 SSMs, 6 UEs; (b) 4 SSMs, 6 UEs; (c) 2 SSMs, 6 UEs.}
            \label{fig:ilcoveragegroup}
        \end{figure}

        To understand the coverage of an SSM in our considered scenario, the cascaded BS-SSM-UE channel $\mathbf{H}_{l,k}$ is analyzed for all the SSMs and with respect to every UE $k$ in the network. Due to the high symmetry in the considered IDS environment, SSMs show similar coverage patterns regardless of their location. Fig.~\ref{fig:solometasurface} shows the coverage condition of a far-away SSM  and a nearby SSM. Approximately 3 rows of UE in front of the reflective side of the SSM are covered for either the far-end SSM case or the near-end SSM case. Thus we consider grouping 1 row of UEs with their nearest 1, 2, or 3 rows of SSMs to form a coverage group. And we index the coverage group with the location row of the UE. Also, in terms of the nearest rows of SSMs, the considered row of UEs should be at the reflective side of the SSMs. As illustrated in Fig.~\ref{fig:ilcoveragegroup}, when we consider grouping 2 rows of SSMs with 1 row of UEs, UEs at row $i$ will be grouped with 4 SSMs that are at row $i$ and row $i-1$ to form coverage group $i$. 

        To determine the size of the coverage group, we consider an over-optimistic case where the self-sustainable constraints of the SSMs are neglected, and all the elements are used to enhance communication. In this case, the SSMs can be assigned to serve all the UEs in their coverage group, in other words, $\alpha_{l,k}=1, \forall l, k$ holds. One upper bound on the SNR under this hypothesis is
        \begin{equation}
            \frac{\left\|\mathbf{h}_k+\mathbf{H}_k\boldsymbol{\phi}_k\right\|^2P}{BN_0}\leq \frac{(\left\|\mathbf{h}_k\|+\|\mathbf{H}_k\right\|_F)^2P}{BN_0}=\hat{\Gamma}_k
        \end{equation}
        
\begin{figure}[tb]
            \centering
            \includegraphics[width=.95\linewidth]{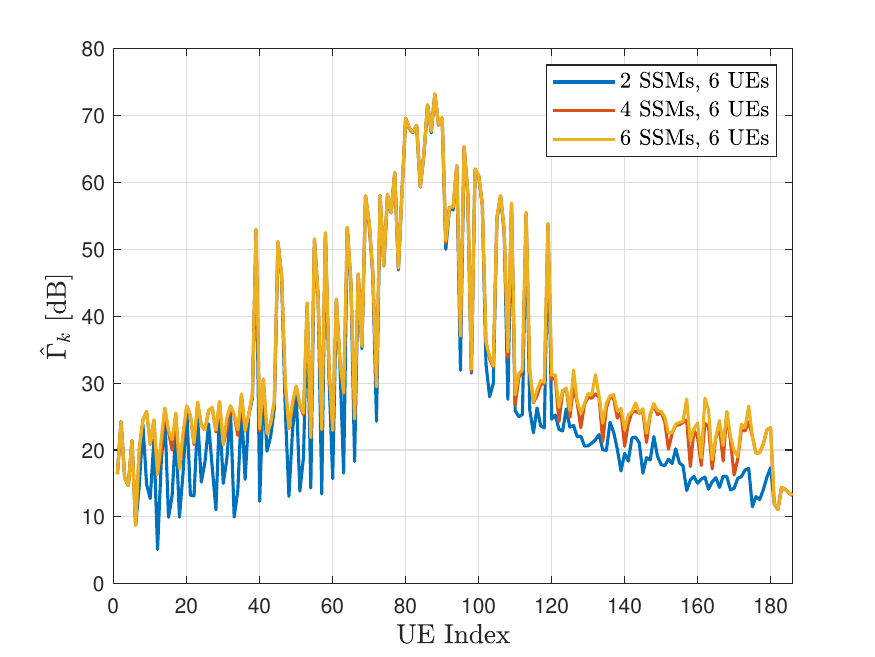}
            \caption{The upper bound of the achievable SNR under different scales of the coverage group.}
            \label{fig:coveragevary}
        \end{figure}

        Fig.~\ref{fig:coveragevary} illustrates $\hat{\Gamma}_k$ when considering grouping 1 row of UEs with 2, 4, and 6 nearest of SSMs. The results indicate that most of the indirect components received by the UE originate from the SSMs located within the same row. While including additional SSMs in the coverage group can yield some benefits, these advantages diminish when more than 4 SSMs are incorporated. Therefore, in this paper, we design the coverage group as comprising 6 UEs with 4 nearest SSMs   (equivalent to group 1 row of UEs and 2 rows of SSMs).


        % To show the optimizing effect of the SSM, the optimized SNR is compared with the SNR obtained with same setup but randomly selected phase-shifts. As shown in Figure~\ref{fig:coveragevary}(b), a gain of about 20 dB can be observed at the edge and in the mid-far region of the cell. And at the area that is close to the BS, there are UEs have reassemble SNR. This is validated in the latter content, which is due to the strong direct component from the BS and the MRT beamforming we considered.

    \subsection{Preset design}

        \begin{figure}[tb]
                \centering
                \includegraphics[width=.95\linewidth]{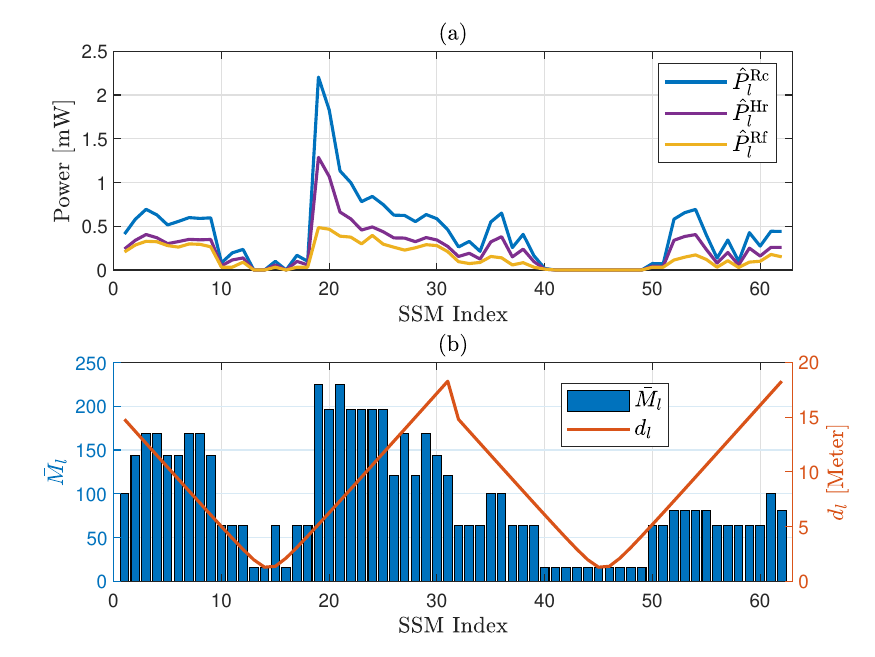}
                \caption{(a) Averaged received power, harvested power, and consumed power under the searched-out reflecting area size with randomized phase-shifts; (b) averaged and rounded down of searched-out reflecting area size and the distance between the SSM and the BS.}
                \label{fig:refsize}
        \end{figure}
        According to \eqref{eq:powerreceived}, due to the MRT beamforming, the self-sustainability of an SSM depends not only on the reflecting area size of a given preset and the distance to the BS, but also on the phase-shifts of all SSMs assigned to serve the target UE. However, these phase-shifts can be only optimized and determined after the reflecting area size of the SSMs is decided. To tackle this issue, we determine the size of the reflecting area by considering an MRT beamformer by randomizing the SSM phase-shifts. The MRT beamformer when serving UE $k$ under the randomized phase-shifts is denoted as $\hat{\mathbf{w}}_k\in\mathbb{C}^{N}$, and is obtained by replacing $\boldsymbol{\phi}_{l,k}$ in \eqref{eq:mrt} with $\hat{\boldsymbol{\phi}}_{l,k}$ where $\hat{\boldsymbol{\phi}}_{l,k}=e^{-j\hat{\boldsymbol{\varphi}}_{l,k}}$, and the entries of $\hat{\boldsymbol{\varphi}}_{l,k}$ are all uniformly distributed in the range of $(0, 2\pi]$. The received power, harvested power, and consumed power of SSM $l$ when serving UE $k$ given $\hat{\mathbf{w}}_k$ are denoted as $\hat{P}_{l,k}^\text{Rc}$, $\hat{P}_{l,k}^\text{Hr}$, and $\hat{P}_{l}^\text{Rf}$ respectively. We search for a largest possible $\bar{M}_l$ for each SSM $l$ from the given set of reflecting area size $\bar{M}_l\in\{
            16 , 64 , 81 , 121 , 144 , 169 , 225 , 256
        \}$ from 10 rounds of Monte-Carlo simulations referring to the utilization criteria of the SSM in the searched situation. We further claim that the utilization criteria of SSM under a given $\bar{M}_l$ refers to the ability to cover all the UEs in the same row as it while maintaining a self-sustainable condition for the MRT beamformer with randomized phase-shifts. We highlight that even though the size of the reflecting area we found is not optimal, it is still a meaningful size that allows the SSM to have a higher chance of being self-sustainable even without optimizing its phase-shifts. For the convenience of evaluating the self-sustainability condition, we denote $\hat{P}^\text{Rc}_{l}$ and $\hat{P}^\text{Hr}_{l}$ as the averaged minimum received and harvested power over all Monte-Carlo simulations, respectively, across various serving UE within all the coverage groups associated with SSM $l$. Fig.~\ref{fig:refsize} illustrates powers and the reflecting area sizes of the SSMs. Fig.~\ref{fig:refsize}(a) shows all the metasurfaces with searched-out reflecting area sizes that can achieve self-sustainability. The relationship between the distance from SSM $l$ to the BS, which is denoted as $d_l$, and the reflecting area size $\bar{M}_l$ is shown in Fig.~\ref{fig:refsize}(b). The results show that the reflecting area size is smaller at the SSMs that are close to the BS, whereas it is larger for SSMs positioned at the cabin's edge and farther away from the BS. The SSM with the largest reflecting area size is situated between the far-end and middle of the cabin. This positioning is attributed to the utilization of the MRT beamforming approach we have taken. The cascaded channel is dominated by the direct channel for UEs that are close to the BS, power will mainly be distributed to the direct channel, thus the insufficient received power leads to a smaller reflecting area of SSM. When the direct channel is not dominant, SSMs receive more power yet the received power decays with the distance to the BS. Therefore, the largest reflecting area appears in between the middle and the edge of the cabin. A detailed analysis of this observation will be presented in Section~\ref{sec:feasibility}. The reflecting area size determined in this section will be used for the later analysis of this work.

\section{Numerical Results} \label{sec:results}

    \begin{figure}
        \centering
        \includegraphics[width=.95\linewidth]{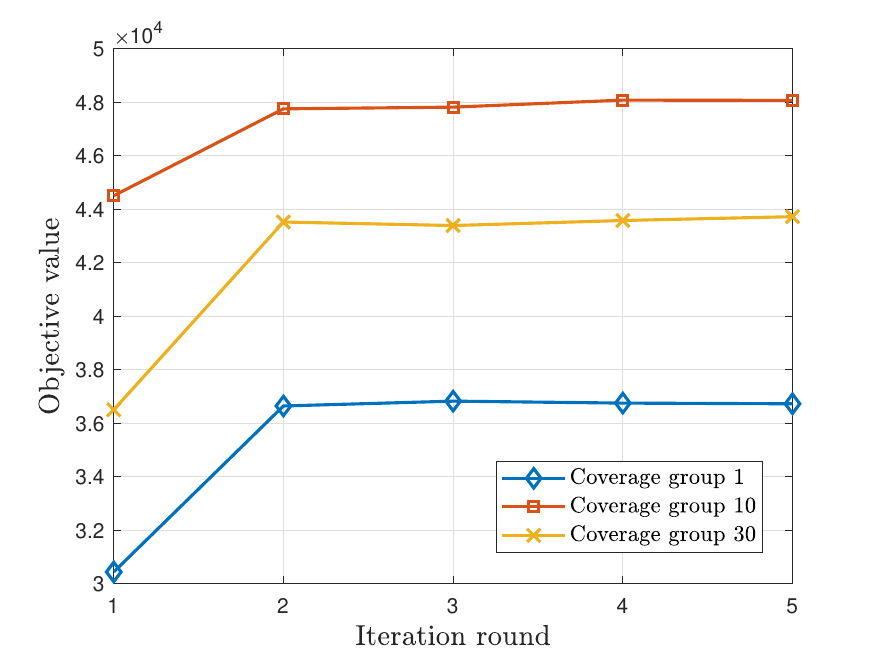}
        \caption{Convergence condition of applying Algorithm~\ref{alg:intraoptimize} concerning the inputs from coverage group 1, 10, and 30.}
        \label{fig:convergence}
    \end{figure}
    Once the channel coefficients have been obtained by RT simulation, the proposed two-stage iterative data rate optimization algorithm can be applied in conjunction with the preset and coverage group design. The power consumption of one reflecting element $P^{\text{Rf}}=2\mu W$~\cite{ntontin2022wireless}. The coefficients featuring power harvesting $q_1$, $q_2$, and $q_3$ are set to be 0.3904, 0.8260, and 0.6823~\cite{le2008efficient}.
    
    We set the maximum number of iteration rounds at $E=5$. Based on our tests, we set the penalty coefficients as $\omega=100$, $\omega'=10^7$, and $\omega''=10^7$ to guarantee that constraints are not violated. Fig.~\ref{fig:convergence} illustrates a snapshot case of the convergence condition of applying Algorithm~\ref{alg:intraoptimize} with respect to the inputs from coverage groups 1, 10, and 30. Comparable convergence conditions are observed regardless of the coverage group, with convergence typically achieved within approximately 2 to 3 rounds of iteration for the algorithm.

    \subsection{Feasibility and necessity of utilizing SSM}\label{sec:feasibility}
    
        \begin{figure}[tb]
            \centering
            \includegraphics[width=.95\linewidth]{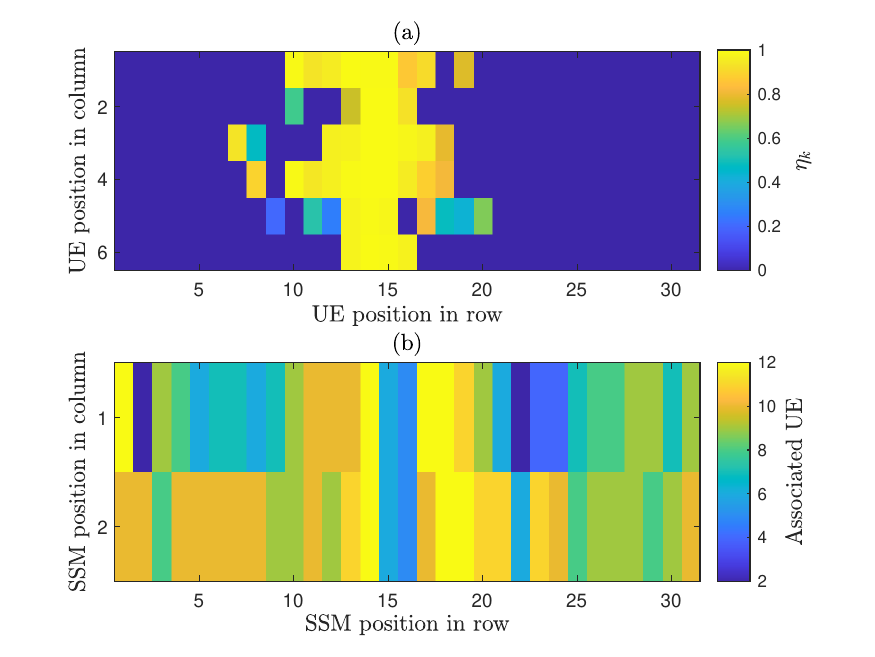}
            \caption{(a) BS contribution for each UE; (b) SSM-UE association condition.}
            \label{fig:association}
        \end{figure}
        Due to propagation attenuation, in some locations, the harvested power may be insufficient to support the operation of the SSM. We define the capability of SSMs to maintain self-sustainability as the \emph{feasibility}. Additionally, SSMs' contributions vary based on their deployment locations when serving UEs directly connected to the BS. We define the capability of SSMs to outperform in terms of contributing to enhancing the UE's SNR over direct connectivity as the \emph{necessity}. To better understand the feasibility and the necessity of utilizing the SSMs at different locations of the cabin, we first evaluate how much gain we obtain by introducing the SSM to the system. To this end, we define a contribution indicator metric $\eta_k\in\mathbb{R}$. We denote the SNR of the UE $k$ considering only the BS with MRT beamforming as $\Gamma_{k}^\text{BS}$, and give the closed-form expression as
        $\Gamma_{k}^\text{BS}=\frac{\|\mathbf{h}_k\|^2P}{BN_0}$.
        We can define the $\eta_k$ as
        \begin{equation}
            \eta_k=\frac{\Gamma_k^\text{BS}}{\Gamma_k},
        \end{equation}
        which is directly proportional to the magnitude of the direct component $\|\mathbf{h}_k\|^2$. As the strength of the direct component received by UE $k$ increases, $\eta_k$ approaches 1. For a UE that does not have a direct connection to the BS, $\eta_k$ will be 0. 
        
        Fig.~\ref{fig:association}(a) shows the $\eta_k$ distribution of each UE in the cabin. It can be seen that almost all UEs that have direct links to the BS have a $\eta_k\geq0.8$, which are distributed in the closest 6 rows to the BS. This indicates in that area, the transmit power is more aligned to the direct channel which leads to a much higher contribution from the direct component rather than via the SSMs. This observation well aligns with the conclusion given in~\cite{9424177}. Fig.~\ref{fig:association}(b) shows the number of associated UEs to a specific SSM. It can be observed that SSMs at the edge of the cabin are associated with more UEs, whereas SSMs located in the middle of the cabin, closer to the BS, are associated with fewer UEs. Combining with the results in Fig.~\ref{fig:association}(a), it can be concluded that SSMs positioned in the middle of the cabin neither provide comparable gains nor are associated with serving more UEs. Consequently, placing SSMs in that area can be a low necessity when designing the SSM-assisted system.

\begin{figure}[tb]
                \centering
                \includegraphics[width=.95\linewidth]{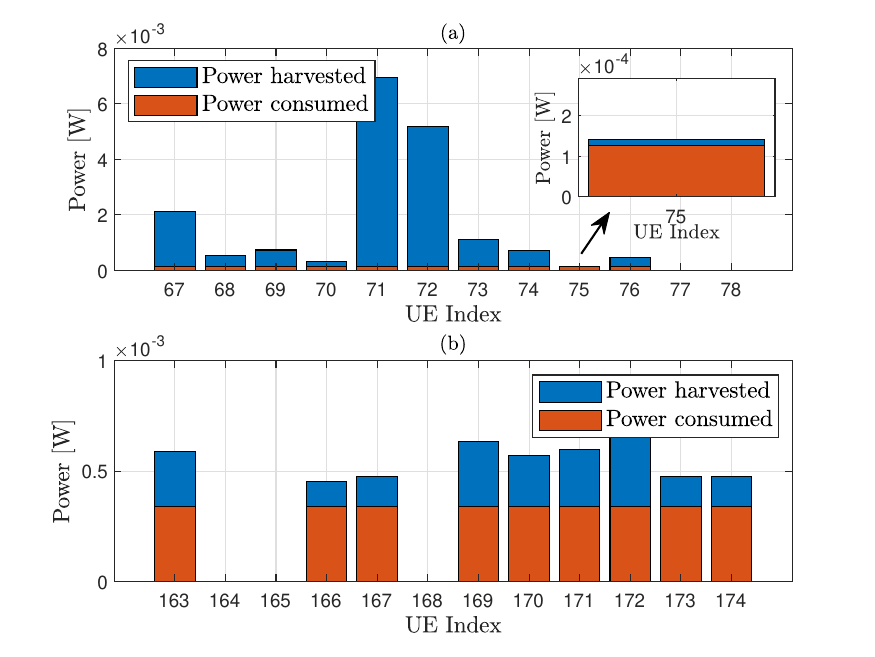}
                \caption{Self-sustainability condition of (a) SSM 12, close to the BS, $\bar{M}_{12}=64$; (b) SSM 29, far from the BS, $\bar{M}_{29}=169$.}
                \label{fig:selfsustainability}
        \end{figure}

        The self-sustainability conditions to the solutions to Algorithm \ref{alg:intraoptimize} are analyzed for the sake of understanding the feasibility of using SSMs. All of the SSMs are observed to satisfy the self-sustainability conditions after the simulations. Fig.~\ref{fig:selfsustainability} illustrates a snapshot case, where we show the power harvested and the power consumed by SSM $12$ and $29$ when they are serving different UEs in their coverage groups. The former is in the middle of the cabin and closer to the BS, and the latter is at the edge of the cabin and far away from the BS.

        From the figure it can be seen that both SSM 12 and SSM 29 harvest more power than they consume for all the assigned UEs, showing that they are self-sustainable. Interestingly, as shown in Fig.~\ref{fig:selfsustainability}(a), despite the shorter distance to the BS, SSM 12 could harvest smaller power than SSM 29. This could also be explained by checking $\eta_k$ given in Fig.~\ref{fig:association}(a). When serving UEs that have direct connections to the BS, most of the transmit power is directed toward the direct channel. As a result, only a small portion of power is captured by the SSMs. Moreover, the SSMs close to the BS are limited by the lowest harvested power and, as a result, cannot support a larger reflecting area, resulting in the harvested power not being sufficiently used in some cases. In the case of SSM 29, all its serving UEs do not have direct connection to the BS, thus power is only allocated to the indirect channel. As shown in Fig.~\ref{fig:selfsustainability}(b), most of the harvested power is used to support its operation regardless of its serving target, suggesting wellness both in the design of the preset and in the use of power.
        
        % However, considering the resulting SNR shown in Figure~\ref{fig:timeallocation}~(a), despite the relatively low contribution from nearby SSMs, UEs close to the BS still achieve higher SNR compared to those without direct connectivity to the BS. This suggests that the MRT beamforming scheme aligns the transmitting beam more towards the direct channel, where the channel condition is significantly better compared to the indirect channel.  Additionally, the power is not directed toward the SSMs, which elucidates why they tend to harvest less power when serving UEs near the BS.

        \begin{figure}[tb]
            \centering
            \includegraphics[width=.95\linewidth]{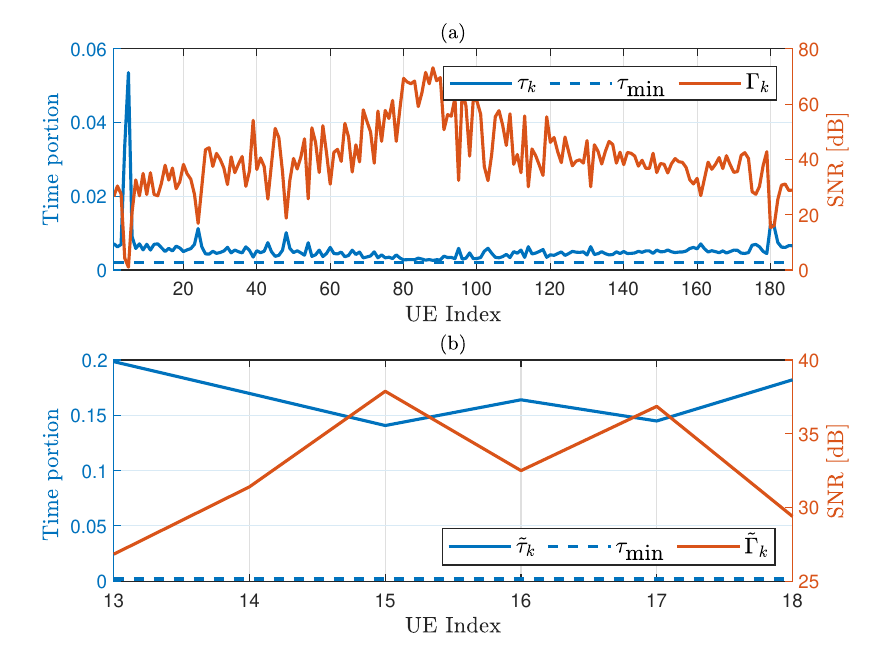}
            \caption{(a) Time allocation and the resulting SNR of the whole cabin; (b) intra-group time allocation and the optimized SNR of the coverage group that consists of UEs 13-18.}
            \label{fig:timeallocation}
        \end{figure}

        Finally, the time resource allocation condition is evaluated, and is illustrated in Fig.~\ref{fig:timeallocation}. Fig.~\ref{fig:timeallocation}(a) shows that all UEs achieve an SNR larger than 0\,dB, which proves the feasibility of employing SSM to guarantee connectivity for the whole cabin. The same figure shows that the resulting SNR decreases roughly as the distance between the UE and BS increases. Moreover, more time resources are allocated to UEs with lower SNR values, while those with better SNRs are allocated with less time resources. UEs near the BS receive the least amount of time resources, yet still more than the minimum allocable portion. This means that the resulting data rate of each UE in the network is expected to be similar, which later gets validated as shown in Fig.~\ref{fig:datarate}. Fig.~\ref{fig:timeallocation}(b) shows the time allocation condition outputted by applying Algorithm~\ref{alg:intraoptimize} concerning one coverage group. Similar to what is shown in Fig.~\ref{fig:timeallocation}(a), the allocated time portion inside the coverage group is also negatively related to the SNR of the UEs.

    \subsection{Achievable data rate}

        \begin{figure}[tb]
            \centering
                \subfloat[]{
                    \label{fig:metasurfacecompare}
                    \includegraphics[trim={70mm 65mm 90mm 60mm},clip,width=.7\linewidth]{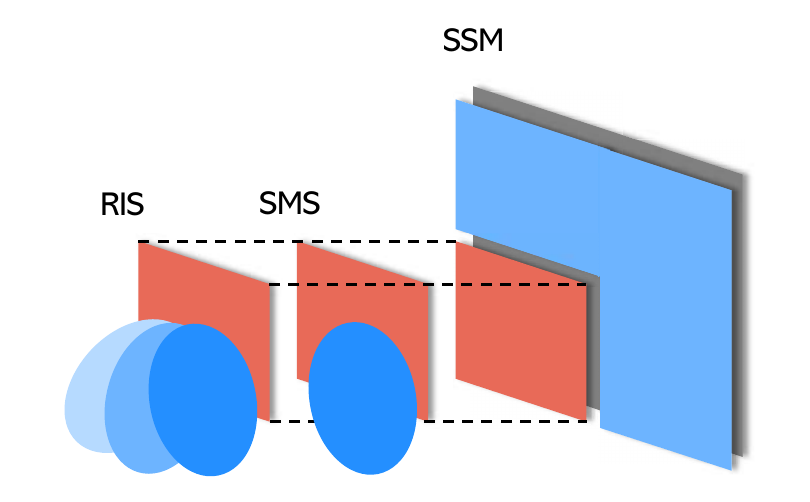}        
                }
                
                \subfloat[]{
                    \label{fig:cellcompare}
                    \includegraphics[trim={80mm 60mm 100mm 50mm},clip,width=.7\linewidth]{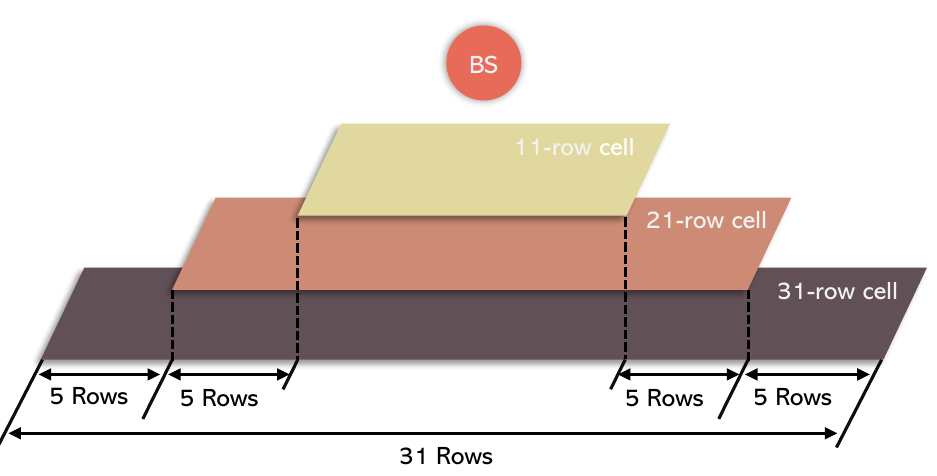}                    
                }
            \caption{Illustrations of metasurface performance test. (a) Setups for different types of metasurface; (b) considered cabins of different sizes and their relative position.}
        \end{figure}
        To determine the placement of SSM, SMS, and RIS within the trade-offs involving metasurface gain, system coverage, and operating cost, we conduct a comparison of the minimum data rate achieved with SSM-assisted, SMS-assisted, and RIS-assisted systems. To fairly compare the performance of SSM with the performance of other types of metasurfaces, we first consider a scenario illustrated in Fig.~\ref{fig:metasurfacecompare}. We consider the size of the RIS and SMS to be equal to the reflective area of the SSM. We also place them geographically in the same position. Considering this, the channel coefficient received by every element within the RIS and the SMS would align with what is received by each reflecting element within the reflecting area of the SSM. Since RIS and SMS are not restricted by self-sustainability, they are utilized referring to the system model mentioned in Fig.~\ref{fig:systemmodel}, where all the RISs/SMSs will jointly serve every UE. The remaining simulation settings are the same for both the RIS and SMS cases. 
        
        Considering the huge attenuation of the mmWave signals, the feasibility of achieving self-sustainability in different sizes of indoor environments for SSM is also expected to be different. It can be expected that when the SSM is barely reaching self-sustainable, the received signal will be prioritized to align with the harvesting channel to ensure its operation, thus less power will be reflected to the UE by the reflective part of the SSM. To explore the influence of self-sustainable feasibility, a performance comparison is conducted across various scenarios employing different types of metasurfaces, considering varying cabin sizes. As illustrated in Fig.~\ref{fig:cellcompare}, three cases, where the cabin consists of UEs and SSMs in 31 rows, 21 rows, and 11 rows, respectively, are considered. In all three cases, the BS is always placed in the center of the cabin.

        \begin{figure}[tb]
            \centering
            \includegraphics[width=.95\linewidth]{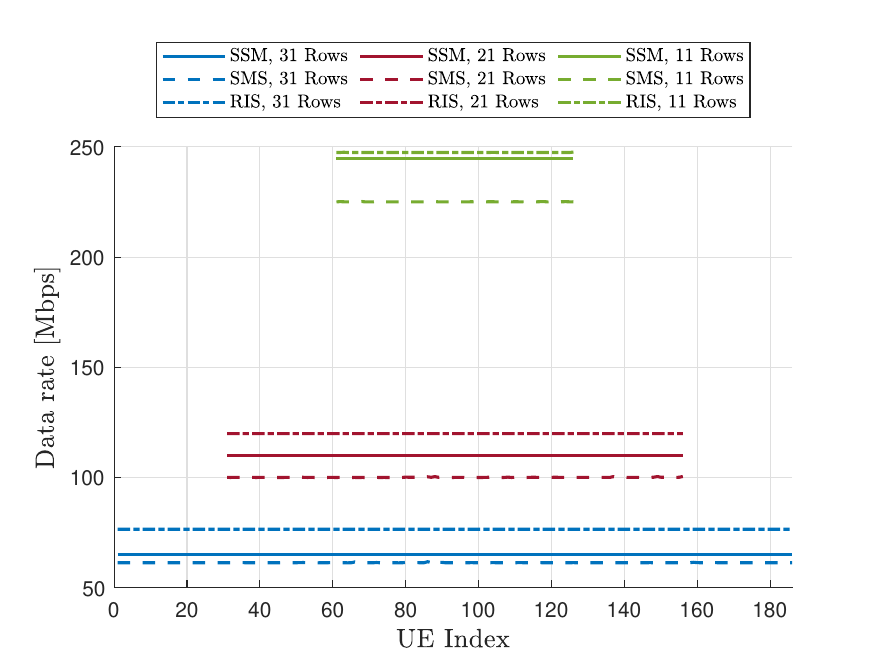}
            \caption{Optimized data rate under the assistance of different types of metasurface and under different cell sizes.}
            \label{fig:datarate}
        \end{figure}
        After obtaining the channel coefficients, the system performance with the assistance of RIS and SMS is evaluated by utilizing the algorithm we proposed in our previous work~\cite{li2024mixed}, where, the minimum data rate is optimized by jointly allocating time resources and selecting the phase-shifts. The results are shown in Fig.~\ref{fig:datarate} along with the system performance we obtained with SSM. %In all the examined scenarios, it is evident that the data rate attained through the utilization of RIS consistently surpasses that achieved by other types of metasurfaces, with SSM and SMS following.
        This observation underscores the trade-off between reconfigurability and operating cost. Under the same channel conditions, RIS outperforms SMS benefiting notably from its reconfigurable nature. The performance gap of that RIS over SMS is $15.2$, $19.8$, and $22.4$\,Mbps on average for each UE when the cabin gets smaller. As a self-sustainable alternative of RIS, SSM, has shown a performance better than SMS due to its reconfigurability, yet a performance worse than RIS due to the extra limitation of self-sustainability. The performance gap of SSM over SMS is $3.6$, $9.9$, and $19.8$\,Mbps for a cabin consisting of 31, 21, and 11 rows, respectively. It is worth noting that in the 11-row cabin case, due to the short distance from the SSMs to the BS, the self-sustainability for the SSMs is much easier to achieve. Since the reflecting areas of the SSMs share the same channel condition as the RISs, the RIS-assisted system performance can be considered as a good upper bound for the SSM. The small gap shows that the simplifications we make do not degrade the performance of the SSM much when the feasibility of achieving self-sustainability is not tight.    
        % However, this does not mean that SSM should always be considered over SMS when operating cost budgets are limited. %Even though the data rate achieved by SSM is higher than what is achieved by SMS in all tested cases, it is still 
          
        % In the 11-row cell case, due to a relatively short propagation distance, the self-sustainability is more feasible to achieve.
        %a higher ratio of metasurfaces in this cell can achieve self-sustainability.
        % Hence, the signal received at the reflective area of the SSM resembles the signal received by the RIS. Consequently, in such instances, SSM performs similarly to RIS. In the 31-row cell case, 
        It is also worth noticing that the data rate gap obtained with SSM over SMS is getting less significant as the cabin size gets larger. With increasing distance to the BS, it becomes more challenging for SSMs to achieve self-sustainability. Thus, the resulting phase-shifts that are applied to SSM will differ from those in the RIS case. In this instance, phase-shifts are selected to affect the transmitting beamforming, which further helps to improve the amount of harvesting power yet losing the metasurface gain SSMs can provide. Thus, as the coverage requirement increases, the performance gap between RIS and SSM widens, while the gap between SSM and SMS narrows. It can be expected that the coverage of SSM stops when it is infeasible to be self-sustainable. In summary, even though SSM shows a better performance than SMS, limited by the self-sustainable constraint, its coverage will be worse than using SMS, which is passive by its nature. Thus, when the system coverage requirement is high, SMS is still a possible solution to enhance the communication quality.
        
\section{Conclusion}\label{sec:conclusion}
    \begin{figure}[tb]
        \centering
        \includegraphics[trim={82mm 35mm 75mm 30mm},clip,width=.6\linewidth]{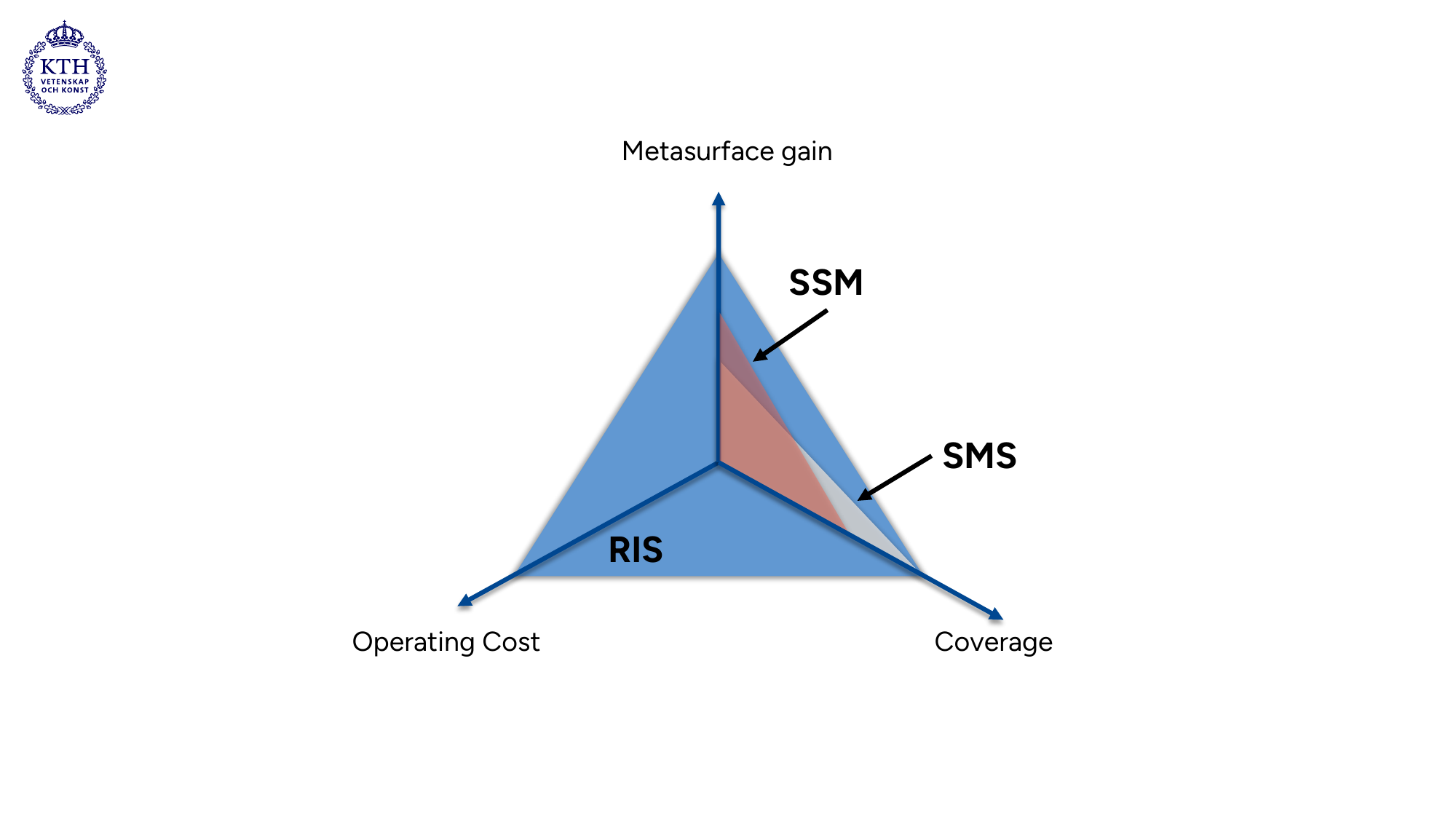}
        \caption{General comparison in terms of operating cost, metasurface gain, and system coverage among SSM, SMS, and RIS.}
        \label{fig:conclusion}
    \end{figure}
    In this paper, in terms of finding a better balance between the metasurface gain, system coverage, and operating cost such as cabling and powering required by the reconfigurability, we investigate the performance of an SSM-assisted system. The metasurface is considered to take the element splitting scheme to achieve self-sustainability, where part of the elements in the metasurface are used only to harvest power from the electromagnetic wave, and the rest part of the elements are used to reflect the signal. To lower the computational complexity of optimizing the SSM-assisted system, certain simplifications are done. Instead of deciding the working mode of each element we consider using the SSM in a preset-based manner, where the whole SSM is divided into the reflecting area and the harvesting area. The preset of each SSM is designed and given beforehand by considering an MRT beamformer in which the phase-shifts of each SSM are randomly selected. Moreover, the SSMs and UEs are associated to form a coverage group based on the coverage of the SSM to further lower the computational complexity. The feasibility and the necessity of using SSMs are evaluated by checking the self-sustainable and the association conditions for SSMs at various locations. Our findings suggest that in areas where UEs benefit from a strong direct channel from the BS, the transmit power tends to align more with this direct channel. Consequently, less power will be harvested by SSMs, and limited by the minimum harvested power employing SSMs in such areas becomes less feasible. Furthermore, for the same reason the contribution of employing SSM is less significant. This indicates a lower necessity for enhancing UEs in those areas with SSM technology. The performance of using RIS, SMS, and SSM is compared and concluded in Fig.~\ref{fig:conclusion}. The results show that SSM brings a performance better than SMS and worse than RIS. The results also show the coverage shortfall of SSM, while trying to use SSM to improve the UEs that are far from the BS, the difficulty for SSM to maintain its self-sustainable condition increases, and this in the end will result in a smaller coverage compared to other types of metasurfaces.

%{\appendices
%\section*{Proof of the First Zonklar Equation}
%Appendix one text goes here.
% You can choose not to have a title for an appendix if you want by leaving the argument blank
%\section*{Proof of the Second Zonklar Equation}
%Appendix two text goes here.}

% \section{References Section}
% You can use a bibliography generated by BibTeX as a .bbl file.
%  BibTeX documentation can be easily obtained at:
%  http://mirror.ctan.org/biblio/bibtex/contrib/doc/
%  The IEEEtran BibTeX style support page is:
%  http://www.michaelshell.org/tex/ieeetran/bibtex/
 
 % argument is your BibTeX string definitions and bibliography database(s)
%\bibliography{IEEEabrv,../bib/paper}
%

\bibliographystyle{IEEEtran}
\bibliography{Main}

\end{document}